\documentclass{arxiv}
\pdfoutput=1
\usepackage{amsmath}
\usepackage{graphicx}
\usepackage{balance}  
\usepackage{times}
\usepackage{caption}
\usepackage{subcaption}
\usepackage{todonotes}
\usepackage[activate={true,nocompatibility},final,tracking=true,factor=1100,stretch=10,shrink=10]{microtype}
\usepackage{array}
\usepackage{url}

\usepackage{enumerate}
\usepackage{listings}

\usepackage{hyphenat} 

\usepackage{algorithmicx}
\usepackage[noend]{algpseudocode}
\usepackage{float}

\usepackage{pifont}
\newcommand{\tick}{\ding{51}} 
\newcommand{\cross}{\ding{53}} 

\newcommand{\mtbase}{MTBase}
\renewcommand{\todo}[2][\inline]{} 

\expandafter\def\expandafter\UrlBreaks\expandafter{\UrlBreaks
  \do\a\do\b\do\c\do\d\do\e\do\f\do\g\do\h\do\i\do\j%
  \do\k\do\l\do\m\do\n\do\o\do\p\do\q\do\r\do\s\do\t%
  \do\u\do\v\do\w\do\x\do\y\do\z\do\A\do\B\do\C\do\D%
  \do\E\do\F\do\G\do\H\do\I\do\J\do\K\do\L\do\M\do\N%
  \do\O\do\P\do\Q\do\R\do\S\do\T\do\U\do\V\do\W\do\X%
  \do\Y\do\Z}

\renewcommand{\footnotesize}{\scriptsize}

\definecolor{mygreen}{rgb}{0,0.6,0}
\definecolor{mygray}{rgb}{0.5,0.5,0.5}
\definecolor{mymauve}{rgb}{0.58,0,0.82}

\newcolumntype{L}[1]{>{\raggedright\let\newline\\\arraybackslash\hspace{0pt}}m{#1}}
\newcolumntype{C}[1]{>{\centering\let\newline\\\arraybackslash\hspace{0pt}}m{#1}}
\newcolumntype{R}[1]{>{\raggedleft\let\newline\\\arraybackslash\hspace{0pt}}m{#1}}

\newtheorem{definition}{Definition}
\newtheorem{cor}{Corollary}

\floatstyle{boxed}

\newfloat{algo}{ht!}{algo}

\floatname{algo}{\linebreak Algorithm}

\begin{document}

\title{\mtbase: Optimizing Cross-Tenant Database Queries}

\numberofauthors{1} 
\author{
\alignauthor Lucas Braun*, Renato Marroqu\'{i}n$^\dag$, Kai-En Tsay$^\dag$, Donald
    Kossmann$^{\dag\sharp}$\\
    \affaddr{\vspace{10pt}* Oracle Corporation, lucas.braun@oracle.com}, work performed
        while at ETH\\
    \affaddr{$^\dag$Systems Group, Department of Computer Science, ETH Zurich,
        \{marenato, tsayk, donaldk\}@ethz.ch}\\
    \affaddr{$^\sharp$Microsoft Corporation, donaldk@microsoft.com}
}


\maketitle

\begin{abstract}
In the last decade, many business applications have moved into the cloud.
In particular, the ``database-as-a-service'' paradigm has become mainstream.
While existing multi-tenant data management systems
focus on
single-tenant query processing, we believe that it is time to rethink how
queries can be processed across multiple tenants in such a way that
we do not only gain more valuable insights, but also at minimal cost.
As we will argue in this paper, standard SQL semantics are insufficient to
process cross-tenant queries in an unambiguous way, which is why existing
systems use other, expensive means like ETL or data integration. We first
propose MTSQL, a set of extensions to standard SQL, which fixes the ambiguity
problem.  Next, we present \mtbase, a query processing middleware that
efficiently processes MTSQL on top of SQL. As we will see, there is a canonical,
provably correct, rewrite algorithm from MTSQL to SQL, which may however result
in poor query execution performance, even on high-performance database products.
We further show that with carefully-designed optimizations, execution times can
be reduced in such ways that the difference to single-tenant queries becomes
marginal.
\end{abstract}
\section{Introduction}
\label{sec:intro}

Indisputably, cloud computing is one of the fastest growing businesses related
to the field of computer science. Cloud providers promise good elasticity, high
availability and a fair pay-as-you-go pricing model to their tenants. Moreover,
corporations are no longer required to rely on on-promise infrastructure which
is typically costly to acquire and maintain. While it is still an open research
question whether and how these good promises can be kept with regard to
databases \cite{das2013elastras, loesing2015architectures}, all the big players,
like Google \cite{krishnan2015google}, Amazon \cite{amazonRds}, Microsoft
\cite{AzureSql} and recently Oracle \cite{oracleCloud}, have launched their own
Database-as-a-Service (DaaS) cloud products. All these products host massive
amounts of clients and are therefore {\em multi-tenant}.



As pointed out by Chong et al.~\cite{chong2006multitenancyspectrum}, the term
{\em multi-tenant database} is ambiguous and can refer to a variety of DaaS
schemes with different degrees of logical data sharing between tenants. On the
other hand, as argued by Aulbach et al.~\cite{aulbach2008multi}, multi-tenant
databases not only differ in the way how they logically share information
between tenants, but also how information is physically separated. We conclude
that the {\em multi-tenancy spectrum} consists of four different schemes: First,
there are DaaS products that offer each tenant her proper database while relying
on physically-shared resources ({\em SR}), like CPU, network and storage.
Examples include {\em SAP HANA} \cite{sap-hana}, {\em SqlVM}
\cite{narasayya2013sqlvm}, {\em RelationalCloud} \cite{mozafari2013dbseer} and
{\em Snowflake} \cite{dageville2016snowflake}. Next, there are systems that
share databases ({\em SD}), but each tenant gets her own set of tables within
such a database, as for in example {\em Azure SQL DB} \cite{das2016daas}.

\begin{figure}[ht!]
    \centering
    \includegraphics[width=\columnwidth]{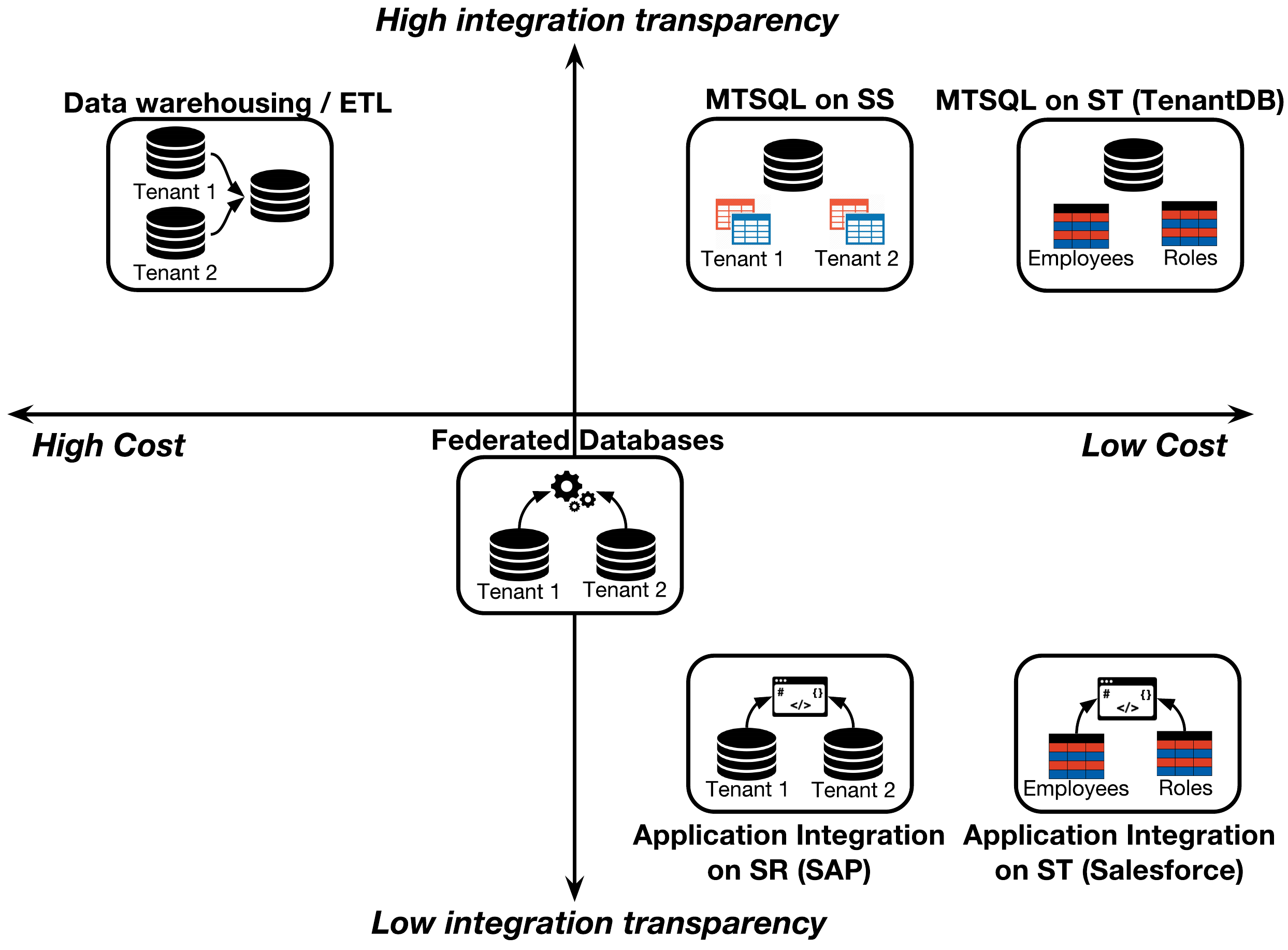}
    \caption{Cross-tenant query processing systems}
    \label{fig:mt-quadrant}
    \vspace{-2mm}
\end{figure}

Finally, there are the two schemes where tenants not only share a database, but
also the table layout (schema). Either, as for example in {\em Apache Phoenix}
\cite{ApachePhoenix}, tenants still have their private tables, but these tables
share the same (logical) schema ({\em SS}), or the data of different tenants is
consolidated into shared tables ({\em ST}) which is hence the layout with the
highest degree of physical and logical sharing. {\em SS} and {\em ST} layouts
are not only used in DaaS, but also in Software-as-a-Service (SaaS) platforms,
as for example in {\em Salesforce} \cite{weissman2009salesforce} and {\em
    FlexScheme} \cite{aulbach2008multi,aulbach2011extensibility}. The main
reason why these systems prefer {\em ST} over {\em SS} is cost
\cite{aulbach2008multi}. Moreover, if the number of tenants exceeds the number
of tables a database can hold ,which is typically a number in the range of ten
thousands, {\em SS} becomes prohibitive.  Conversely, {\em ST} databases can
easily accommodate hundred thousands to even millions of tenants.

An important feature of {\em multi-tenant databases}, which we believe did
not yet get the attention it deserves, is {\em cross-tenant query
    processing}.
One compelling use case is health care where many providers and insurances use
the same integrated SaaS application. If the providers would agree to query
their joint datasets of (properly anonymized) patient data with scientific
institutions, this could enable medical research to advance much faster because
the data can be queried as soon as it gets in.

\todo[inline]{copy-pasted from LB thesis, maybe not needed as moving related
    work upfront, would also solve the problem... [done]}

This paper looks into {\em cross-tenant query processing} within the scope of
{\em SS} and {\em ST} databases, thereby optimizing a very specific sub-problem
of {\em data integration}\index{data integration} (DI). DI, in a broad sense, is
about finding schema and data mappings between the original schemas of different
data sources and a target schema specified by the client
application~\cite{niswonger1999clio,fagin2009clio,raman2001potter}. As such, DI
techniques are applicable to the entire spectrum of multi-tenant databases
because even if tenants use different schemas or databases, these techniques can
identify correlations and hence extract useful information. Our work embraces
and builds on top of the latest DI work.
More concretely, we optimize conversion functions similar to those used in DI by
thoroughly analyzing and exploiting their algebraic properties.
In addition, instead of translating data into a specific client format (and
update periodically), we convert it to any required client format efficiently
and {\em just-in-time}.
\todo[inline]{end-copy [done]}

There are several existing approaches to {\em cross-tenant query processing}
which are summarized in Figure~\ref{fig:mt-quadrant}: The first approach is {\em data
    warehousing} \cite{kimball2002data} where data is {\em extracted} from
several {\em data sources} (tenant databases/tables), {\em transformed} into
one common format and finally {\em loaded} into a new database where it can be
queried by the client. This approach has high integration transparency in the
sense that once the data is loaded, it is in the right format as required by the
client and she can ask any query she wants. Moreover, as all data is in a single
place, queries can be optimized. On the down-side of this approach, as argued by
\cite{braun2015analytics,neumann2015fast,arulraj2016bridging}, are costs in
terms of both, developing and maintaining such {\em ETL} pipelines and
maintaining a separate copy of the data. Another disadvantage is {\em data
    staleness} in the presence of frequent updates.

{\em Federated Databases} \cite{levy1998information,haas2002data} reduce some of
these costs by integrating data {\em on demand}, i.e. there is no copying.
However, maintenance costs are still significant as for every new data source a
new integrator/wrapper has to be developed.  As data resides in different
places (and different formats), queries can only be optimized to a very small
extent (if at all), which is why the degree of integration transparency is
considered sub-optimal.  Finally, systems like {\em SAP HANA} \cite{sap-hana}
and {\em Salesforce} \cite{weissman2009salesforce}, which are mainly tailored
towards single-tenant queries, offer some degree of {\em cross-tenant query
    processing}, but only through their application logic, which means that the
set of queries that can be asked is limited.

The reason why none of these approaches tries to use SQL for {\em cross-tenant
    query processing} is that it is ambiguous. Consider, for instance the {\em
    ST} datbase in Figure~\ref{fig:ex:running}, which we are going to use as a
running example throught the paper:
As soon as we want to query the joint dataset of tenants 0 and 1 and, for
instance, join {\tt Employees} with {\tt Roles}, joining on {\tt role\_id} alone
is not enough as this would also join {\tt Patrick} with {\tt researcher} and
{\tt Ed} with {\tt professor}, which is clear nonsense. The obvious solution is
to add the tenant-ID {\tt ttid} to the join predicate. On the other hand,
joining the {\tt Employees} table with itself on {\tt E1.age = E2.age} does not
require {\tt ttid} to be present in the join predicate because it actually makes
sense to include results like {\tt (Alice, Ed)} as they are indeed the same age.
As {\tt ttid} is an attribute invisible to the client, plain SQL has no way to
distinguish the two cases, the one where {\tt ttid} has to be included in the
join and the one where it does not. Another challenge arises from the fact that
different tenants might store their employees' salaries in different currencies.
If this is the case, computing the average salary across all tenants clearly
involves some value conversions that should, ideally, happen without the client
noticing or even worrying about.

This paper presents {\em MTSQL} as a solution to all these ambiguity problems.
MTSQL extends SQL with additional semantics specifically-suited for {\em
    cross-tenant query processing}. It enables high integration transparency
because any client, with any desired data format, can ask any query at any time.
Moreover, as data resides in a single database ({\em SS} or {\em ST}), queries
can be aggressively optimized with respect to both, standard SQL semantics and
additional MTSQL semantics. As MTSQL adopts the single-database layout, it is
also very cost-effective, especially if used on top of {\em ST}. Moreover, data
conversion only happens as needed, which perfectly fits the cloud's {\em
    pay-as-you-go} cost model.
Specifically, the paper makes the following contributions:

\begin{itemize}
    \item It defines the syntax and semantics of {\em MTSQL}, a query language
        that extends SQL with additional semantics suitable for cross-tenant
        query processing.
    \item It presents the design and implementation of {\em \mtbase}, a database
        middleware that executes {\em MTSQL} on top of any {\em ST} database.
    \item It studies MTSQL-specific optimizations for query execution in \mtbase.
    \item It extends the well-known TPC-H benchmark to run and evaluate {\em
            MTSQL} workloads.
    \item It evaluates the performance and the implementation correctness of
        {\em \mtbase} with this benchmark.
\end{itemize}

The rest of this paper is organized as follows: Section~\ref{sec:mtsql} defines {\em
    MTSQL} while Section~\ref{sec:mtbase} gives an overview on {\em \mtbase}.
Section~\ref{sec:optimizations} discusses the MTSQL-specific optimizations, which are
validated in Section~\ref{sec:experiments} using the benchmark presented in
Section~\ref{sec:mt-bench}. Section~\ref{sec:related-work} shortly summarizes lines of
related work while the paper is concluded in Section~\ref{sec:conclusion}.

\begin{figure}[ht!]
    \centering
    \begin{subtable}[b]{\columnwidth}
        \tabcolsep 2pt
        \centering{\scriptsize
        \begin{tabular}{|c|c|l|c|c|r|c|}
            \hline
            {\bf E\_ttid} & {\bf E\_emp\_id} & {\bf E\_name} & {\bf E\_role\_id} & {\bf
                E\_reg\_id} & {\bf E\_salary} & {\bf E\_age}\\ \hline  
            \hline
            0 & 0 & Patrick & 1 & 3 & 50K   & 30 \\ \hline
            0 & 1 & John    & 0 & 3 & 70K   & 28 \\ \hline
            0 & 2 & Alice   & 2 & 3 & 150K  & 46 \\ \hline
            \hline
            1 & 0 & Allan   & 1 & 2 & 80K   & 25 \\ \hline
            1 & 1 & Nancy   & 2 & 4 & 200K  & 72 \\ \hline
            1 & 2 & Ed      & 0 & 4 & 1M    & 46 \\ \hline
        \end{tabular}
        }
        \captionsetup{justification=centering}
        \caption*{\scriptsize {\bf Employees (tenant-specific)},\\{\tt E\_salary} of tenant 0 in USD,
        {\tt E\_salary} of tenant 1 in EUR}
        \label{tab:ex:emps}
        \vspace{3mm}
    \end{subtable}
    
    \begin{subtable}[b]{.48\columnwidth}
        \tabcolsep 2pt
        \centering{\scriptsize
        \begin{tabular}{|c|c|l|}
            \hline
            {\bf R\_ttid} & {\bf R\_role\_id} & {\bf R\_name}\\ \hline 
            \hline
            0 & 0 & phD stud.   \\ \hline
            0 & 1 & postdoc     \\ \hline
            0 & 2 & professor   \\ \hline 
            \hline
            1 & 0 & intern      \\ \hline
            1 & 1 & researcher  \\ \hline
            1 & 2 & executive   \\ \hline
        \end{tabular}
        }
        \caption*{\scriptsize \bf Roles (tenant-specific)}
        \label{tab:ex:roles}
    \end{subtable}%
    \begin{subtable}[b]{.48\columnwidth}
        \tabcolsep 2pt
        \centering{\scriptsize
        \begin{tabular}{|c|l|}
            \hline
            {\bf Re\_reg\_id} & {\bf Re\_name}\\ \hline 
            \hline
            0 & AFRICA \\ \hline
            1 & ASIA \\ \hline
            2 & AUSTRALIA \\ \hline
            3 & EUROPE \\ \hline
            4 & N-AMERICA \\ \hline
            5 & S-AMERICA \\ \hline
        \end{tabular}
        }
        \caption*{\scriptsize \bf Regions (global)}
        \label{tab:ex:regions}
    \end{subtable}
    \captionsetup{justification=centering}
    \caption{MTSQL database in {\em Basic Layout} ({\em ST}),
        \\$ttid$s not visible to clients}
    \label{fig:ex:running}
\end{figure}

\begin{figure}[ht!]
    \centering
    \begin{subtable}[c]{\columnwidth}
        \tabcolsep 2pt
        \centering{\scriptsize
        \begin{tabular}{|c|l|c|c|r|c|}
            \hline
            {\bf E\_emp\_id} & {\bf E\_name} & {\bf E\_role\_id} & {\bf
                E\_reg\_id} & {\bf E\_salary} & {\bf E\_age}\\ \hline  
            \hline
            0 & Patrick & 1 & 3 & 50K   & 30 \\ \hline
            1 & John    & 0 & 3 & 70K   & 28 \\ \hline
            2 & Alice   & 2 & 3 & 150K  & 46 \\ \hline
        \end{tabular}
        }
        \caption*{\scriptsize {\bf Employees\_0 (private)}, {\tt E\_salary} in
            USD}
        \label{tab:private:ex:emps1}
        \vspace{3mm}
    \end{subtable}
    \begin{subtable}[c]{\columnwidth}
        \tabcolsep 2pt
        \centering{\scriptsize
        \begin{tabular}{|c|l|c|c|r|c|}
            \hline
            {\bf E\_emp\_id} & {\bf E\_name} & {\bf E\_role\_id} & {\bf
                E\_reg\_id} & {\bf E\_salary} & {\bf E\_age}\\ \hline  
            \hline
            0 & Allan   & 1 & 2 & 80K   & 25 \\ \hline
            1 & Nancy   & 2 & 4 & 200K  & 72 \\ \hline
            2 & Ed      & 0 & 4 & 1M    & 46 \\ \hline
        \end{tabular}
        }
        \caption*{\scriptsize {\bf Employees\_1 (private)}, {\tt E\_salary} in
            EUR}
        \label{tab:private:ex:emps2}
        \vspace{3mm}
    \end{subtable}
    
    \begin{subtable}[c]{.5\columnwidth}
        \centering
        \begin{subtable}[c]{0.85\columnwidth}
            \tabcolsep 2pt
            \raggedright{\scriptsize
            \begin{tabular}{|c|l|}
                \hline
                {\bf R\_role\_id} & {\bf R\_name}\\ \hline 
                \hline
                0 & phD stud.   \\ \hline
                1 & postdoc     \\ \hline
                2 & professor   \\ \hline 
            \end{tabular}
            }
            \caption*{\raggedright \scriptsize \bf Roles\_0 (private)}
            \label{tab:private:ex:roles1}
            \vspace{3mm}
        \end{subtable}
        \\
        \begin{subtable}[c]{.85\columnwidth}
            \tabcolsep 2pt
            \raggedright{\scriptsize
            \begin{tabular}{|c|l|}
                \hline
                {\bf R\_role\_id} & {\bf R\_name}\\ \hline 
                \hline
                0 & intern      \\ \hline
                1 & researcher  \\ \hline
                2 & executive   \\ \hline
            \end{tabular}
            }
            \caption*{\raggedright \scriptsize \bf Roles\_1 (private)}
            \label{tab:private:ex:roles2}
        \end{subtable}%
    \end{subtable}%
    \begin{subtable}[c]{.45\columnwidth}
        \tabcolsep 2pt
        \centering{\scriptsize
        \begin{tabular}{|c|l|}
            \hline
            {\bf Re\_reg\_id} & {\bf Re\_name}\\ \hline 
            \hline
            0 & AFRICA \\ \hline
            1 & ASIA \\ \hline
            2 & AUSTRALIA \\ \hline
            3 & EUROPE \\ \hline
            4 & N-AMERICA \\ \hline
            5 & S-AMERICA \\ \hline
        \end{tabular}
        }
        \caption*{\scriptsize \bf Regions (global)}
        \label{tab:private:ex:regions}
    \end{subtable}
    \caption{MTSQL database in {\em Private Table Layout} ({\em SS})}
    \label{fig:private:ex:running}
\end{figure}

\section{MTSQL}
\label{sec:mtsql}

In order to model the specific aspects of {\em cross-tenant query processing} in
{\em multi-tenant databases}, we developed {\em MTSQL}, which will be described
in this section. MTSQL extends SQL in two ways: First, it extends the SQL
interface with two additional parameters, $C$ and $D$.  $C$ is the tenant ID (or
{\em ttid} for short) of the client who submits a statement and hence determines
the format in which the result must be presented.  The data set, $D$, is a set
of {\em ttids} that refer to the tenants whose data the client wants to query.
Secondly, MTSQL extends the syntax and semantics of SQL, as well as its Data
Definition Language (DDL), Data Manipulation Language (DML) and Data Control
Language (DCL, consists of {\tt GRANT} and {\tt REVOKE} statements).

As mentioned in the introduction, there are several ways how a multi-tenant
database can be laid out: Figure~\ref{fig:ex:running} shows an example of the {\em ST}
scheme, also referred to as {\em basic layout} in related work
\cite{aulbach2008multi} where tenants's data is consolidated using the same
tables. Meanwhile, Figure~\ref{fig:private:ex:running} illustrated the {\em SS}
scheme, also referred to as {\em private table layout}, where every tenant has
her own set of tables. In that scheme, {\em data ownership} is defines as part
of the table name while in {\em ST}, records are explicitly annotated with the
{\em ttid} of their {\em data owner}, using an extra meta column in the table
which is invisible to the client.

As these two approaches are semantically equivalent, the MTSQL semantics that we
are about to define apply to both. In the case of the {\em SS}, applying a
statement $s$ with respect to $D$ simply means to apply $s$ to the logical union
of all private tables owned by a tenant in $D$. In {\em SS}, $s$ is applied to
tables filtered according to $D$. In order to keep the presentation simple, the
rest of this paper assumes an {\em ST} scheme, but sometimes defines semantics
with respect to {\em SS} if that makes the presentation easier to understand.

\subsection{MTSQL API}
\label{sec:mtsql:api}

MTSQL needs a way to incorporate the additional parameters $C$ and $D$. As $C$
is the {\em ttid} of the tenant that issues a statement, we assume it is
implicitly given by the SQL connection string. {\em ttids} are not only used for
identification and access control, but also for data ownership (as shown in
Figure~\ref{fig:private:ex:running}). While this paper uses integers for simplicity
reasons, {\em ttids} can have any data type, in particular they can also be
database user names.

$D$ is defined using the MTSQL-specific {\tt SCOPE} runtime parameter on the SQL
connection. This parameter can be set in two different ways:
Either, as shown in
Listing~\ref{lst:simple-scope-exp}, as {\em simple scope} with an {\tt IN} list stating
the set of {\em ttids} that should be queried,
or as in Listing~\ref{lst:complex-scope-exp}, as
a sub-query with a {\tt FROM} and a {\tt WHERE} clause ({\em complex scope}).
The semantics of the latter is that every tenant that owns at least one record
in one of the tables mentioned in the {\tt FROM} clause that satisfies the {\tt
    WHERE} clause is part of $D$.
The SCOPE variable defaults to $\{C\}$, which means that by default a client
processes only her own data. Defining a simple scope with an empty {\tt IN}
list, on the other hand, makes $D$ include all the tenants present in the
database.

Making $C$ and $D$ part of the connection allowed for a clear separation between
the end users of MTSQL (for which {\tt ttids} do not make much sense and hence
remain invisible) and administrators/programmers that manage connections (and
are aware of {\tt ttids}).
\todo[inline]{Instead of having a footnote, I wrote that last sentence about the
    visibility of ttids... can someone double-check, please?}

\begin{lstlisting}[numbers=none,caption={Simple SCOPE expression using IN},
    label={lst:simple-scope-exp}, language=SQL]
SET SCOPE = "IN (1,3,42)";
\end{lstlisting}

\begin{lstlisting}[numbers=none,caption={Complex SCOPE expression with
        sub-query}, label={lst:complex-scope-exp}, language=SQL]
SET SCOPE = "FROM Employees WHERE E_salary > 180K";
\end{lstlisting}

\subsection{Data Definition Language}

DDL statements are issued by a special role called the {\em data modeller}. In a
multi-tenant application, this would be the SaaS provider (e.g. a Salesforce
administrator) or the provider of a specific application. However, the data
modeller can delegate this privilege to any tenant she trusts using a {\tt
    GRANT} statement, as will be described in \S~\ref{sec:mtsql:dcl}. 

There are two types of tables in MTSQL: tables that contain common knowledge
shared by everybody (like {\tt Regions}) and those that contain data of a
specific tenant (i.e. {\tt Employees} and {\tt Roles}). More formally, we define
the {\em table generality} of {\tt Regions} as {\em global} and the one of all
other tables as {\em tenant-specific}. In order to process queries across
tenants, MTSQL needs a way to distinguish whether an attribute is {\em
    comparable} (can be directly compared against attribute values of other
tenants), {\em convertible} (can be compared against attribute values of other
tenants after applying a well-defined {\em conversion function}) or {\em
    tenant-specific} (it does semantically not make sense to compare against
attribute values of other tenants). An overview of these types of {\em attribute
    comparability}, together with examples from Figure~\ref{fig:ex:running}, is shown
in Figure~\ref{fig:attribute-comparabilty}.

\begin{table}[ht!]
    \centering{\scriptsize
    \begin{tabular}{|C {2.0cm}|C {2.5cm}|C {2.5cm}|}
        \hline
        {\bf type} & {\bf description} & {\bf examples} \\
        \hline \hline
        comparable & can be directly compared to and aggregated with other
        values & E\_region\_id, E\_age, Re\_name, R\_region\_id, R\_name \\
        \hline
        convertible & other values need to be converted to the format of the
        current tenant before comparison or aggregation & E\_salary \\
        \hline
        tenant-specific & values of different tenants cannot be compared with
        each other & E\_role\_id, R.role\_id \\
        \hline
    \end{tabular}
    \caption{Overview on attribute comparability in MTSQL}
    \label{fig:attribute-comparabilty}
    }
\end{table}

\subsubsection{CREATE TABLE Statement}
\label{sec:mtsql:keywords}

The MTSQL-specific keywords for creating (or altering) tables are {\tt GLOBAL,
    SPECIFIC, COMPARABLE} and {\tt CONVERTIBLE}. An example of how they can
be used is shown in Listing~\ref{lst:create-table}. Note that {\tt SPECIFIC} can be
used for tables and attributes. Moreover, using these keywords is optional as we
define that tables are global by default, attributes of tenant-specific tables
default to {\em tenant-specific} and those of global tables to {\em
    comparable}.\footnote{In fact, global tables, as they are shared between all
    tenants, can only have comparable attributes anyway.}

\newpage

\begin{lstlisting}[caption={Exemplary MTSQL CREATE TABLE statement, MT-specific
    keywords marked in bold},
    label={lst:create-table}, language=SQL]
CREATE TABLE Employees SPECIFIC (
  E_emp_id  INTEGER     NOT NULL SPECIFIC,
  E_name    VARCHAR(25) NOT NULL COMPARABLE,
  E_role_id INTEGER     NOT NULL SPECIFIC,
  E_reg_id  INTEGER     NOT NULL COMPARABLE,
  E_salary  VARCHAR(17) NOT NULL CONVERTIBLE @currencyToUniversal @currencyFromUniversal,
  E_age     INTEGER     NOT NULL COMPARABLE,
  CONSTRAINT pk_emp PRIMARY KEY (E_emp_id),
  CONSTRAINT fk_emp FOREIGN KEY (E_role_id) REFERENCES Roles (R_role_id)
);
\end{lstlisting}

\subsubsection{Conversion Functions}
\label{sec:mtsql:conv-functions}

Cross-tenant query processing requires the ability to execute comparison
predicates on {\em comparable} and {\em convertible attribute}.  While
comparable attributes can be directly compared to each other, convertible
attributes, as their name indicates, have to be converted first, using
conversion functions. Each tenant has a pair of conversion functions for each
attribute to translate from and to a well-defined universal format. More
formally, a {\em conversion function pair} is defined as follows:

\begin{definition}\label{def:conversion-funcs}
    {\normalfont 
    $(toUniversal : X \times T \rightarrow X, from$-\linebreak $Universal : X
    \times T \rightarrow X)$ is a valid MTSQL conversion function pair for
    attribute $A$, where $T$ is the set of tenants in the database and $X$ is
    the domain of $A$, if and only if:
    \begin{enumerate}[(i)]
        \item\label{def:universal-format} There exists a {\em universal format}
            for attribute $A$:\footnote{$image(f)$ denotes the mathematical
                image, i.e. the range of function $f$.}\\
            $image(toUniversal(\cdot, t_1)) = image(toUniversal(\cdot, t_2))$\\
            $= \ldots = image(toUniversal(\cdot, t_{|T|}))$
        \item For every tenant $t \in T$, the partial functions
            $toUniversal(\cdot, t)$ and $fromUniversal(\cdot, t)$ are
            bijective functions.
        \item $fromUniversal$ is the inverse of $toUniversal$: $\forall t \in
            T,\linebreak x \in X: fromUniversal(toUniversal(x,t), t) = x$
    \end{enumerate}
    }
\end{definition}

\todo[inline]{adjust the look and feel of the equations... [done]}

These three properties imply the following two corollaries that we are going to
need later in this paper:

\begin{cor}\label{def:equiv-preserving}
    {\normalfont
    $toUniversal$ and $fromUniversal$ are equality preserving: $\forall t \in T:
    toUniversal(x,t) = toUniversal(y,t) \Leftrightarrow x = y \Leftrightarrow
    fromUniversal(x,t) = fromUniversal(y,t)$
    }
\end{cor}

\begin{cor}
    {\normalfont
    Values from any tenant $t_i$ can be converted into the representation of any
    other tenant $t_j$ by first applying \linebreak $toUniversal(\cdot, t_i)$,
    followed by $fromUniversal(\cdot, t_j)$ while \linebreak
    equality is preserved:
    \vspace{-5pt}
    \begin{align*}
    \forall t_i, t_j \in T: x = y \Leftrightarrow &
    fromUniversal(toUniversal(x,t_i),t_j) \\
    = & fromUniversal(toUniversal(y,t_i),t_j)
    \end{align*}
    }
\end{cor}

The reason why we opted for a two-step conversion through universal format is
that it allows each tenant $t_i$ to define her share of the conversion function
pair, i.e. $toUniversal(\cdot, t_i)$ and $from$-\linebreak $Universal(\cdot,
t_i)$, individually without the need of a central authority. Moreover, this
design greatly reduces the overall number of partial conversion functions as we
need at most $2\cdot|T|$ partial function definitions, compared to $|T|^2$
functions in the case where we would define a direct conversion for every pair
of tenants.

Listings~\ref{lst:pToUniv} and \ref{lst:pFromUniv} show an example of such a
conversion function pair. These functions are used to convert phone numbers with
different prefixes, like ``+'', ``00'' or any other specific county exit
code\footnote{The country exit code is a sequence of digits that you have to
    dial in order to inform the telco system that you want to call a number
    abroad. A full list of country exit codes can be found on
    \url{http://www.howtocallabroad.com/codes.html}.}, and the universal format
is a phone number without prefix.
In this example, converting phone numbers simply means to lookup the tenant's
prefix and then either prepend or remove it, depending whether we convert from
or to the universal format. Note that the exemplary code also contains the
keyword {\tt IMMUTABLE} to state that for a specific input the function always
returns the same output, which is an important hint for the query optimizer.
While this keyword is PostgreSQL-specific, some other vendors, but by far not
all, offer a similar syntax.

\begin{lstlisting}[language=SQL, firstline=38, lastline=40, caption={Converting a
    phone number to universal form (without
    prefix), PostgreSQL syntax}, label={lst:pToUniv}]
CREATE FUNCTION phoneToUniversal (VARCHAR(17), INTEGER) RETURNS VARCHAR(17)
    AS 'SELECT SUBSTRING($1, CHAR_LENGTH(PT_prefix)+1) FROM Tenant, PhoneTransform WHERE T_tenant_key = $2 AND T_phone_prefix_key = PT_phone_prefix_key;'
LANGUAGE SQL IMMUTABLE;
\end{lstlisting}

\begin{lstlisting}[language=SQL, firstline=42, lastline=44, caption={Converting to
    a specific phone number format, PostgreSQL syntax},
    label={lst:pFromUniv}]
CREATE FUNCTION phoneFromUniversal (VARCHAR(17), INTEGER) RETURNS VARCHAR(17)
    AS 'SELECT CONCAT(PT_prefix, $1) FROM Tenant, PhoneTransform WHERE T_tenant_key = $2 AND T_phone_prefix_key = PT_phone_prefix_key;'
LANGUAGE SQL IMMUTABLE;
\end{lstlisting}

It is important to mention that the {\em equality-preserving} property as
mentioned in Corollary~\ref{def:equiv-preserving} is a minimal requirement for
conversion functions to make sense in terms of producing coherent query results
among different clients.  There are, however conversion functions that exhibit
additional properties, for example:

\begin{itemize}
    \item order-preserving with respect to tenant $t$:\\
        $x < y \Leftrightarrow toUniversal(x,t) < toUniversal(y,t)$
    \item homomorphic with respect to tenant $t$ and function $h$: \\
        $toUniversal(h(x_1, x_2,...),t) =$\\
        $h(toUniversal(x_1, t), toUniversal(x_2, t),...)$
\end{itemize}

We will call a conversion function pair {\em fully-order-preserving} if
$toUniversal$ and $fromUniversal$ are order-preserving with respect to all
tenants. Consequently, a conversion function pair can also be {\em
    fully-{\em h}-preserving}.

Listings~\ref{lst:cToUniv} and \ref{lst:cFromUniv} show an exemplary conversion
function pair used to convert currencies (with USD as universal format). These
functions are not only equality-preserving, but also fully-SUM-preserving: as
the currency conversion is nothing but a multiplication with a constant
factor\footnote{We are aware of the fact that currency conversion is not at all
    constant, but depends on rapidly changing exchange rates.  However, we want
    to keep the examples as simple as possible in order to illustrate the
    underlying concepts. However, the general ideas of this paper also apply to
    temporal databases.} from {\tt CurrencyTransform}, it does not matter in
which format we sum up individual values (as long as they all have that same
format). As we will see, such special properties of conversion functions are
another crucial ingredient for query optimization.

\newpage

\begin{lstlisting}[language=SQL, firstline=30, lastline=32, caption={Converting a
    currency to universal form (USD), PostgreSQL
    syntax},label={lst:cToUniv}]
CREATE FUNCTION currencyToUniversal (DECIMAL(15,2), INTEGER) RETURNS DECIMAL(15,2)
    AS 'SELECT CT_to_universal*$1 FROM Tenant, CurrencyTransform WHERE T_tenant_key = $2 AND T_currency_key = CT_currency_key;'
LANGUAGE SQL IMMUTABLE;
\end{lstlisting}

\begin{lstlisting}[language=SQL, firstline=34, lastline=36, caption={Converting
    from USD to a specific currency, PostgreSQL
    syntax},label={lst:cFromUniv}]
CREATE FUNCTION currencyFromUniversal (DECIMAL(15,2), INTEGER) RETURNS DECIMAL(15,2)
    AS 'SELECT CT_from_universal*$1 FROM Tenant, CurrencyTransform WHERE T_tenant_key = $2 AND T_currency_key = CT_currency_key;'
LANGUAGE SQL IMMUTABLE;
\end{lstlisting}

The conversion function examples shown in Listings \ref{lst:pToUniv} to
\ref{lst:cFromUniv} assume the existence of tables holding additional conversion 
information ({\tt CurrencyTransmform} and {\tt PhoneTransform}) as well as a 
table with references into these tables (named {\tt Tenants} table). The way how a tenant can
define her portion of the conversion functions is then simply to choose a
specific currency and phone format as part of an initial setup procedure.
However, this is only one possible implementation. MTSQL does not make any
assumptions or restrictions on the implementation of conversion function pairs
themselves, as long as they satisfy the properties given in
Definition~{\ref{def:conversion-funcs}}.

MTSQL is not the first work that talks about conversion functions. In fact,
there is an entire line of work that deals with data integration and in
particular with schema mapping techniques
\cite{niswonger1999clio,fagin2009clio,aulbach2008multi}. These works mention and
take into account conversion functions, like for example a multiplication or a
division by a constant. More complex conversion functions, including
regular-expression-based substitutions and other arithmetic operations, can be
found in {\em Potter's Wheel} \cite{raman2001potter} where {\em conversion} is
referred to as {\em value translation}. All these different conversion functions
can potentially also be used in MTSQL which is, to the best of our knowledge,
the first work that formally defines and categorizes conversion functions
according to their properties.

\subsubsection{Integrity Constraints}
\label{sec:mtsql:integrity-constraints}

MTSQL allows for {\em global} integrity constraints that every tenant has to
adhere to (with respect to the entirety of her data) as well as {\em
    tenant-specific} integrity constraints (that tenants can additionally impose
on their own data). An example of a {\em global} referential integrity
constraint is shown in the end of Listing~\ref{lst:create-table}. This
constraint means that for every tenant, for each entry of {\tt E\_role\_id}, a
corresponding entry {\tt R\_role\_id} has to exist in {\tt Roles} and must be
owned by that same tenant. Consider for example employee {\em John}
with\linebreak {\tt R\_role\_id} 0.  The constraint implies that their must be a
{\em role} 0 owned by tenant 0, which in that case is {\em PhD student}. If the
constraint were only {\em tenant-specific} for tenant 1, John would not link to
roles and {\tt E\_role\_id} 0 would just be an arbitrary numerical value. In
order to differentiate {\em global} from {\em tenant-specific} constraints, the
scope is used\footnote{Remembering that an empty IN list refers all tenants,
    this is exactly what is used to indicate a global constraint. Additionally,
    all constraints created as part of a {\tt CREATE TABLE} statement are global
    as well.}.
\todo[inline]{adjusted this paragraph to take into account the modified scope
    definition --> please proof-read!}

\subsubsection{Other DDL Statements}
\label{sec:mtsql:ddl-others}

{\tt CREATE VIEW} statements look the same as in plain SQL.  As for the other
DDL statements, anyone with the necessary privilege can define global
views on {\em global} and {\em tenant-specific} tables. Tenants are allowed to
create their own, tenant-specific views (using the default scope).
The selected data has to be presented in
universal format if it is a {\em global} view and in the {\em tenant-specific} format
otherwise.
{\tt DROP VIEW}, {\tt DROP TABLE} and {\tt ALTER TABLE} work the same way as in
plain SQL.

\subsection{Data Control Language}
\label{sec:mtsql:dcl}

Let us have a look at the MTSQL {\tt GRANT} statement:

\begin{lstlisting}[caption={MTSQL GRANT syntax}, numbers=none, language=SQL]
GRANT <privileges> ON <database|table> TO <ttid>;
\end{lstlisting}

As in plain SQL, this grants some set of access privileges (READ, INSERT, UPDATE
and/or DELETE) to the tenant identified by $ttid$.  In the context of MTSQL,
however, this means that the privileges are granted with respect to $C$.
Consider the following statement:

\begin{lstlisting}[caption={Example of an MTSQL GRANT statement}, numbers=none,
    language=SQL]
GRANT READ ON Employees TO 42;
\end{lstlisting}

In the {\tt private table layout}, if $C$ is 0, then this would grant tenant 42
read access to {\tt Employees\_0}, but if $C$ is 1, tenant 42 would get read
access to {\tt Employees\_1} instead. If a grant statement grants to $ALL$, then
the grant semantics also depend on $D$, more concretely if $D=\{7,11,15\}$ the
privileges would be granted to tenants 7, 11 and 15.

By default, a new tenant that joins an MTSQL system is granted the following
privileges: READ access to global tables, READ, INSERT, UPDATE, DELETE,
GRANT and REVOKE on his own instances of tenant-specific tables. In our example,
this means that a new tenant 111 can read and modify data in {\tt
    Employees\_111} and {\tt Roles\_111}. Next, a tenant can start asking around
to get privileges on other tenants' tables or also on global tables.
The {\tt REVOKE} statement, as in plain SQL, simply revokes privileges that were granted
with {\tt GRANT}.

\subsection{Query Language}
\label{sec:mtsql:query-language}

Just as in FlexScheme \cite{aulbach2008multi, aulbach2011extensibility}, queries
themselves are written in plain SQL and have to be filtered according to $D$.
Whereas in FlexScheme $D$ always equals $\{C\}$ (a tenant can only query her own
data), MTSQL allows cross-tenant query processing, which means that the data set
can include other tenants than $C$ and can in particular be bigger than one.  As
mentioned in the introduction, this creates some new challenges that have to be
handled with special care.

\begin{figure*}[ht!]
    \centering
    \includegraphics[width=.75\textwidth]{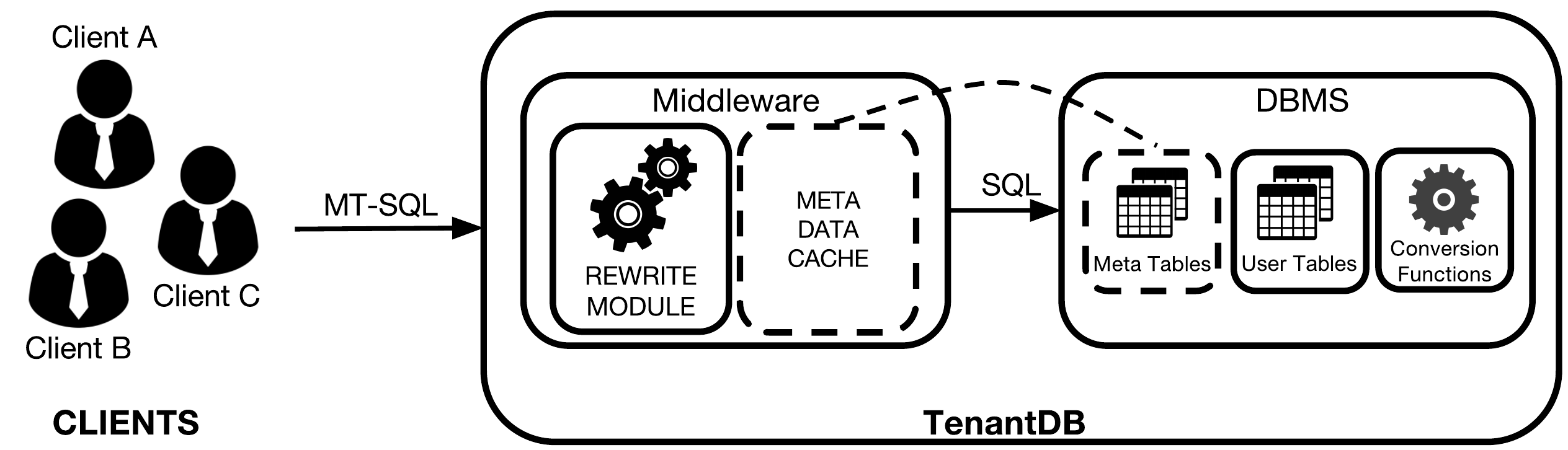}
    \caption{\mtbase~architecture}
    \label{fig:impl:mtbase-arch}
\end{figure*}

\subsubsection{Client Presentation}

As soon as tenants can query other tenants' data, the MTSQL engine has to be
make sure to deliver results in the proper format. For instance, looking again
at Figure~\ref{fig:ex:running}, if tenant 0 queries the average salary of all
employees of tenant 1, then this should be presented in USD because tenant 0
stores her own data in USD and expects other data to be in USD as well.
Consequently, if tenant 1 would ask that same query, the result would be
returned as is, namely in EUR.

\subsubsection{Comparisons}
\label{sec:mtsql:query-language:comparisons}

Consider a join of {\tt Roles} and {\tt Employees} on {\tt reg\_id}. As long as
the dataset size is only one, such a join query has the same semantics as in
plain SQL (or FlexScheme). However, as soon as tenant 1, for instance, asks this
query with $D=\{0,1\}$, the join has to take the {\em ttids} into account. The
reason for this is that {\tt reg\_id} is a {\em tenant-specific} attribute and
should hence only be joined within the same tenant in order to prevent
semantically wrong results like John being an intern (although tenant 0 does
not have such a role) or Nancy being a professor (despite the fact that tenant 1
only has roles {\em intern}, {\em researcher} and {\em executive}).

Comparison or join predicates containing {\em comparable} and {\em convertible}
attributes, on the other hand, just have to make sure that all data is brought
into universal format before being compared. For instance, if tenant 0 wants to
get the list of all employees (of both tenants) that earn more than 100K USD,
all employee salaries have to be converted to USD before executing the
comparison.

Finally, MTSQL does not allow to compare {\em tenant-specific} with other
attributes. For instance, we see no way how it could make sense to compare {\tt
    E\_role\_id} to something like {\tt E\_age} or \linebreak {\tt E\_salary}.

\subsection{Data Manipulation Language}
\label{sec:mtsql:dml}

\todo[inline]{Rewrote this paragraph and removed unnecessary restriction that
    insert statements can only have scope size 1. Please proof-read!}
MTSQL DML works the same way as in FlexScheme
\cite{aulbach2008multi,aulbach2011extensibility} if $D=\{C\}$.
Otherwise, if $D \neq \{C\}$, the semantics of a DML statement are defined such
that it is applied to each tenant in $D$ separately. Constants, {\tt WHERE}
clauses and sub-queries are interpreted with respect to $C$, exactly the same
way as for queries (c.f.~\S~\ref{sec:mtsql:query-language}). This implies that
executing {\tt UPDATE} or {\tt INSERT} statements might involve value conversion
to the proper tenant format(s).

\section{\mtbase}
\label{sec:mtbase}

Based on the concepts described in the previous section, we implemented \mtbase,
an open-source MTSQL engine~\cite{mtbase}. As shown in
Figure~\ref{fig:impl:mtbase-arch}, the basic building block of \mtbase~is an
MTSQL-to-SQL translation middleware sitting between a traditional DBMS and the
client. In fact, as it communicates to the DBMS (and to the client) by the means
of pure SQL, \mtbase\@ works in conjunction with any off-the-shelve DBMS. For
performance reasons, the proxy maintains a cache of MT-specific meta data, which
is persisted in the DBMS along with the actual user data.  Conversion functions
are implemented as UDFs that might involve additional meta tables, both of which
are also persisted in the DBMS. \mtbase~implements the {\em basic data layout},
which means that {\em data ownership} is implemented as an additional (meta)
$ttid$ column in each {\em tenant-specific} table as illustrated in
Figure~\ref{fig:ex:running}). There are some dedicated meta tables: {\tt Tenant}
stores each tenant's privileges and conversion information and {\tt Schema}
stores information about table and attribute comparability. Additional meta
tables can (but do not have to) be used to implement conversion function pairs,
as for example {\tt CurrencyTransform} and {\tt PhoneTransform} shown in
Listings \ref{lst:pToUniv} to \ref{lst:cFromUniv}.

While the rewrite module was implemented in Haskell and compiled with GHC \cite{GHC},
the connection handling and the meta data cache maintenance was written in
Python (and run with the Python2 interpreter)~\cite{Python2}. Haskell is handy
because we can make full use of pattern matching and additive data types to
implement the rewrite algorithm in a quick and easy-to-verify way, but any other
functional language, like e.g. Scala~\cite{ScalaLang}, would also do the job.
Likewise, there is nothing fundamental in using Python, any other framework that
has a good-enough abstraction of SQL connections, e.g. JDBC~\cite{JDBC}, could
be used.

\todo[inline]{adjusted the following paragraph to take into account the modified
    scope definition --> please proof-read!}
Upon opening a connection at the middleware, the client's $ttid$, $C$, is
derived from the connection string and used throughout the entire lifetime of
that connection. Whenever a client sends a MTSQL statement $s$, first if the
current scope is complex, a SQL query $q_s$  is derived from this scope and
evaluated at the DBMS in order to determine the relevant dataset $D$. After
that, $D$ is compared against privileges of $C$ in the {\tt Tenant} table and
$ttid$s in $D$ without the corresponding privilege are pruned, resulting in
$D'$.  Next, $C$, $D'$ and $s$ are input into the rewrite algorithm which
produces a rewritten SQL statement $s'$ which is then sent to the DBMS before
relaying the result back to the client. Note that in order to guarantee
correctness in the presence of updates, $q_s$ and $s'$ have to be executed
within the same transaction and with a consistency level at least {\em
    repeatable-read} \cite{berenson1995isolationlevels} (even if the client does
not impose any transactional guarantees). If $s$ is a DDL statement, the
middleware also updates the MT-Specific meta information in the DBMS and the
cache.

The rest of this section explains the MTSQL-to-SQL rewrite algorithm in its
canonical form and proves its correctness with respect to
\S~\ref{sec:mtsql:query-language}, while Section~\ref{sec:optimizations} shows
how to optimize the rewritten queries such that they can be run on the DBMS with
reasonable performance.

\subsection{Canonical Query Rewrite Algorithm}
\label{sec:mtbase:query-rewrite}

Our proposed canonical MTSQL-to-SQL rewrite algorithm works top-down, starting
with the outer-most SQL query and recursively rewriting sub-queries as they come
along.  For each sub-query, the SQL clauses are rewritten one-by-one.  The
algorithm makes sure that for each sub-query the following invariant holds: the
result of the sub-query is filtered according to $D'$ and presented in the
format required by $C$.

The pseudo code of the general rewrite algorithm for rewriting a (sub-)query is
shown in Algorithm~\ref{algo:rewrite-general}. Note that {\tt FROM}, {\tt GROUP
    BY}, {\tt ORDER BY} and {\tt HAVING} clause can be rewritten without any
additional context while {\tt SELECT} and {\tt WHERE} need the whole query as an
input because they might need to check the {\tt FROM} for additional
information, for instance they must know to which original tables certain
attributes belong.

\todo[inline]{describe exact input and output [done]}
\setlength{\textfloatsep}{5pt}
\begin{algo}
\begin{algorithmic}[1]
\State {\bf Input:}\hspace{3.2mm}$C$: $ttid$, $D$: set of $ttid$s, $Q$: MTSQL query
\State {\bf Output:} SQL query
\Function{RewriteQuery}{$C, D, Q$}
\State $new$-$select \gets $ rewriteSelect($C$,$D$,$Q$)
\State $new$-$from \gets $ rewriteFrom($C$,$D$,$Q$.from())
\State $new$-$where \gets $ rewriteWhere($C$,$D$,$Q$)
\State $new$-$group$-$by \gets $ rewriteGroupBy($C$,$D$,$Q$.groupBy())
\State $new$-$order$-$by \gets $ rewriteOrderBy($C$,$D$,$Q$.orderBy())
\State $new$-$having \gets $ rewriteHaving($C$,$D$,$Q$.having())
\State \Return new Query ($new$-$select$, $new$-$from$, $new$-$where$,
$new$-$group$-$by$, $new$-$order$-$by$, $new$-$having$)
\EndFunction
\end{algorithmic}
\caption{Canonical Query Rewrite Algorithm}
\label{algo:rewrite-general}
\end{algo}

In the following, we will look at the rewrite functions for the different SQL
clauses. Because of space constraints, we only provide the high-level ideas and
illustrate them with suitable minimal examples. However, we strongly encourage
the interested reader to check-out the Haskell code~\cite{mtrewrite} which in
fact almost reads like a mathematical definition of the rewrite algorithm.


\begin{lstlisting}[caption={Examples for Rewriting SELECT clause},
    label={lst:rewrite-select}, language=SQL]
-- Rewriting a simple select expression:
SELECT E_salary FROM Employees; -->
SELECT currencyFromUniversal(currencyToUniversal(E_salary, ttid), C) as salary FROM Employees;
-- Rewriting an aggregated select expression
SELECT AVG(E_salary) as avg_sal FROM Employees; -->
SELECT AVG(currencyFromUniversal(currencyToUniversal(E_salary, ttid), C)) as avg_sal FROM Employees;
-- Rewriting star expression, hiding irrelevant info
SELECT * FROM Employees; -->
SELECT E_name, E_reg_id, E_salary, E_age FROM Employees;
\end{lstlisting}

\noindent{\bf SELECT} \hspace{2mm}
The rewritten {\tt SELECT} clause has to present every attribute $a$ in $C$'s
format, which, if $a$ is convertible, is achieved by two calls to the conversion
function pair of $a$ as can be seen in the examples of
Listing~\ref{lst:rewrite-select} where \texttt{-\,->} simply denotes rewriting. If $a$ is
part of compound expression (as in line 6), it has to be converted before the
functions (in that case $AVG$) are applied.  Note that in order to make a
potential super-query work correctly, we also rename the result of the
conversion, either by the new name that it got anyway (as in line 6) or by the
name that it had before (as in line 3).  Rewriting a star expression (line 9) in
the uppermost query also needs special attention, in order not to provide the
client with confusing information, like $ttids$ which should stay invisible.
\begin{lstlisting}[caption={Examples for Rewriting WHERE clause},
    label={lst:rewrite-where}, language=SQL]
-- Comparison with a constant:
.. FROM Employees WHERE E_salary > 50K  -->
.. WHERE currencyFromUniversal(currencyToUniversal(E_salary,ttid),C) > 50K) ..
-- General comparison:
.. FROM Employees E1, Employees E2 WHERE E1.E_salary > E2.E_salary -->
.. WHERE currencyFromUniversal(currencyToUniversal(E1.E_salary,E1.ttid),C) > currencyFromUniversal(currencyToUniversal(E1.E_salary,E1.ttid),C) ..
-- Extend with predicate on ttid
.. FROM Employees, Roles WHERE E_role_id = R_role_id -->
.. FROM Employees, Roles WHERE E_role_id = R_role_id AND Employees.ttid = Roles.ttid ..
-- Adding D-filters for D' = {3,7}
.. FROM Employees E, Roles R .. -->
.. WHERE E.ttid IN (3,7) AND R.ttid IN (3,7) ..
\end{lstlisting}

\noindent{\bf WHERE} \hspace{2mm}
There are essentially three steps that the algorithm has to perform in order to
create a correctly rewritten {\tt WHERE} clause (as shown in
Listing~\ref{lst:rewrite-where}). First, conversion functions have to be added to each
convertible attribute in each predicate in order make sure that comparisons are
executed in the correct (client) format (lines 2 to 6).  This happens the same
way as for a {\tt SELECT} clause. Notably, all constants are always in $C$'s
format because it is $C$ who asks the query. Second, for every predicate
involving two or more {\tt tenant-specific} attributes, additional predicates on
$ttid$ have to be added (line 9), unless if the attributes are part of the same
table, which means they are owned by the same tenant anyway. Predicates that
contain {\tt tenant-specific} together with other attributes cause the entire
query to be rejected as was required
in~\S~\ref{sec:mtsql:query-language:comparisons}. Last, but not least, for every
base table in the {\tt FROM} clause, a so-called {D-filter} has to be added to
the {\tt WHERE} clause (line 12). This filter makes sure that only the relevant
data (data that is owned by a tenant in $D'$) gets processed.

\vspace{3.5mm}\noindent{\bf FROM} \hspace{2mm}
All tables referred by the {\tt FROM} clause are either base tables or temporary
tables derived from a sub-query. Rewriting the {\tt FROM} clause simply means to
call the rewrite algorithm on each referenced sub-query as shown in
Algorithm~\ref{algo:rewrite-from}. A {\tt FROM} table might also contain a {\tt
    JOIN} of two tables (sub-queries). In that case, the two sub-queries are
rewritten and then the join predicate is rewritten in the exact same way like
any {\tt WHERE}.

Notably, this algorithm preserves the desired invariant for (sub-) queries: the
result of each sub-query is in client format and filtered according to $D'$,
and, due to the rewrite of the {\tt SELECT} and the {\tt WHERE} clause of the
current query, base tables, as well as joins, are also presented in client
format and filtered by $D$. We conclude that the result of the current query
therefore also preserves the invariant.

\todo[inline]{describe input and output [done]}
\begin{algo}
\begin{algorithmic}[1]
\State {\bf Input:}\hspace{3.2mm}$C$: $ttid$, $D$: set of $ttid$s,
\State\hspace{11.5mm}$FromClause$:
MTSQL FROM clause
\State {\bf Output:} SQL FROM clause
\Function{RewriteFrom}{$C, D, FromClause$}
\State $res \gets $ extractBaseTables $(FromClause)$
\ForAll {$q \in $ extractSubQueries $(FromClause)$}
\State $res \gets res \cup \{$ rewriteQuery $(C, D, q)\}$
\EndFor
\ForAll {$(q_1,q_2,cond) \in $ extractJoins $(FromClause)$}
    \State $q_1' \gets $ rewriteQuery $(C, D, q_1)$
    \State $q_2' \gets $ rewriteQuery $(C, D, q_2)$
    \State $cond' \gets $ rewriteWhere $(C, D, cond)$
    \State $res \gets res \cup \{$ createJoin $(q_1', q_2', cond'))\}$
\EndFor
\Return $res$
\EndFunction
\end{algorithmic}
\caption{Rewrite Algorithm for {\tt FROM} clause}
\label{algo:rewrite-from}
\end{algo}

\noindent{\bf GROUP-BY, ORDER-BY and HAVING} \hspace{2mm}
{\tt HAVING} and\linebreak{\tt GROUP-BY} clauses are basically rewritten the
same way like the expressions in the {\tt SELECT} clause. Some DBMSs might throw
a warning stating that grouping by a comparable attribute $a$ is ambiguous
because the way we rewrite $a$ in the WHERE clause and rename it back to $a$, we
could actually group by the original or by the converted attribute $a$. However,
the SQL standard clearly says that in such a case, the result should be grouped
by the outer-more expression, which is exactly what we need. {\tt ORDER-BY}
clauses need not be rewritten at all.

\todo[inline]{I think that this paragraph can stay as is even with the new scope
definition. Proof-read, please!}
\vspace{3.5mm}\noindent{\bf SET SCOPE} \hspace{2mm}
Simple scopes do not have to be rewritten at all. The {\tt FROM} and {\tt WHERE}
clause of a complex scope are rewritten the same way as in a sub-query. In order
to make it a valid SQL query, the rewrite algorithm adds a {\tt SELECT} clause
that projects on the respective $ttid$s as shown in Listing~\ref{lst:rewriting-scope}.

\begin{lstlisting}[caption={Rewriting a complex SCOPE expression},
    label={lst:rewriting-scope}, language=SQL]
SET SCOPE = "FROM Employees WHERE E_salary > 180K"; -->
SELECT ttid FROM Employees WHERE currencyFromUniversal(currencyToUniversal(E_salary,ttid),C) > 180K;
\end{lstlisting}


\todo[inline]{This proof might be a bit weak... Please proof-read!}

\newpage
\subsection{Algorithm Correctness}

\begin{proof}
We prove the correctness of the canonical rewrite algorithm with respect to
\S~\ref{sec:mtsql:query-language} by induction over the composable structure of
SQL queries and by showing that the desired invariant (the result of each
sub-query is filtered according to $D'$ and presented in the format required by
$C$) holds: First, as a base, we state that adding the D-filters in the {\tt
    WHERE} clause and transforming the {\tt SELECT} clause to client format for
every base table in each lowest-level sub-query ensures that the invariant
holds. Next, as an induction step, we state that the way how we rewrite the {\tt
    FROM} clause, as it was described earlier, preserves that property.  The
top-most SQL query is nothing but a composition of sub-queries (and base tables)
for which the invariant holds. This means that the invariant holds for the
entire query, which is hence guaranteed to deliver the correct result.
\end{proof}

\todo[inline]{moving some examples of DDL and DML to appendix? [done]}

\subsection{Rewriting DDL and DML Statements}
Rewriting DDL and DML statements is very similar to rewriting queries, in fact,
predicates are rewritten in exactly the same way. The remaining questions are
how to rewrite {\em tenant-specific} referential integrity constraints (using
check constraints) and how to apply DML statements to a dataset $D \neq \{C\}$
(by executing the proper value transformations separately for each client).
While the semantics and the intuition how to implement them should be clear, we
refer to Appendix~\ref{app:more-rewriting} for further examples and explanations.

\section{Optimizations}
\label{sec:optimizations}

As we have seen, there is a canonical rewrite algorithm that correctly rewrites
MTSQL to SQL. However, we will show in Section~\ref{sec:experiments} that the
rewritten queries often execute very slowly on the underlying DBMS. The main
reason for this is that the pure rewritten queries call two conversion functions
on every transformable attribute of every record that is processed, which is
extremely expensive. Luckily, the execution costs can be reduced dramatically
when applying the optimization passes that we describe in this section.  As we
assume the underlying DBMS to optimize query execution anyway, we focus on
optimizations that a DBMS query optimizer cannot do (because it needs
MT-specific context) or does not do (because an optimization is not frequent
enough outside the context of \mtbase). We differentiate between {\em semantic
    optimizations}, which are always applied because they never make a query
slower and {\em cost-based} optimizations which are only applied if the
predicted costs are smaller than in the original query.

\begin{lstlisting}[caption={Examples for trivial semantic optimizations},
    label={lst:trivial-opts}, language=SQL]
-- dropping D-filter if D is the empty scope:
SELECT E_age FROM Employees WHERE E_ttid IN (1,2); -->
SELECT E_age FROM Employees;
-- dropping ttid from join predicate if |D| = 1:
SELECT E_age, R_name FROM Employees, Roles WHERE E_role_id = R_role_id AND E_ttid = R_ttid AND E_ttid IN (2) AND R_ttid IN (2); -->
SELECT E_age, R_name FROM Employees, Roles WHERE E_role_id = R_role_id AND E_ttid IN (2) AND R_ttid IN (2);
-- dropping conversion functions if D = {C}:
SELECT currencyFromUniversal(currencyToUniversal(E_salary, E_ttid),0) AS E_salary FROM Employees; -->
SELECT E_salary FROM Employees;
\end{lstlisting}

\subsection{Trivial Semantic Optimizations}

There are a couple of special cases for $C$ and $D$ that allow to save
conversion function calls, join predicates and/or D-filters. First, if $D$
includes all tenants, that means that we want to query all data and hence
D-filters are no longer required as shown in line 3 of
Listing\ref{lst:trivial-opts}.  Second, as shown in line 6, if $|D|=1$, we know
that all data is from the same tenant, which means that including {\tt ttid} in
the join predicate is no longer necessary. Last, if we know that a client
queries her own data, i.e. $D=\{C\}$ corresponds to the default scope, we know
that even convertible attributes are already in the correct format and can hence
remove the conversion function calls (line 9).

\subsection{Other Semantic Optimizations}

There are a couple of other semantic optimizations that can be applied to
rewritten queries. While {\em client presentation push-up} and {conversion
    push-up} minimize the number of conversions by delaying conversion to the
latest possible moment, {\em aggregation distribution} takes into account
specific properties of conversion functions (as mentioned in
\S~\ref{sec:mtsql:conv-functions}). If conversion functions are UDFs written in
SQL it is also possible to inline them.  This typically gives queries an
additional speed up.

\subsubsection{Client Presentation and Conversion Push-Up}
\label{sec:optimizations:push-up}

As conversion function pairs are equality-preserving, it is possible in some
cases to defer conversions to later, for example to the outermost query in the
case of nested queries. While {\em client presentation push-up} converts
everything to universal format and defers conversion to client format to the
outermost {\tt SELECT} clause, {\em conversion push-up} pushes this idea even
more by also delaying the conversion to universal format as much as possible.
Both optimizations are beneficial if the delaying of conversions allows the
query execution engine to evaluate other (less expensive) predicates first. This
means that, once the data has to be converted, it is already more filtered and
therefore the overall number of (expensive) conversion function calls becomes
smaller (or, in the worst case, stays the same). Naturally, if we delay
conversion, this also means that we have to propagate the necessary {\em ttids}
to the outer-more queries and keep track of the current data format.

\begin{lstlisting}[caption={Example for client presentation push-up},
    label={lst:client-push-up}, language=SQL]
-- before optimization
SELECT Dom.name1, Dom.sal1 as sal, COUNT(*) as cnt FROM (
  SELECT E1.name as name1, currencyFromUniversal(currencyToUniversal(E1.E_salary, E1.E_ttid), C) as sal1
  FROM Employees E1, Employees E2
  WHERE currencyFromUniversal(currencyToUniversal(E1.E_salary, E1.E_ttid), C) >
  currencyFromUniversal(currencyToUniversal(E2.E_salary, E2.E_ttid), C)
) as Dom GROUP BY Dom.name1, sal, cnt ORDER BY cnt;
-- after optmimization
SELECT Dom.name1, currencyFromUniversal(Dom.sal1, C) as sal, COUNT(*) as cnt FROM (
  SELECT E1.name as name1, currencyToUniversal(E1.E_salary, E1.E_ttid) as sal1
  FROM Employees E1, Employees E2
  WHERE currencyToUniversal(E1.E_salary, E1.E_ttid) > currencyToUniversal(E2.E_salary, E2.E_ttid)
) as Dom GROUP BY Dom.name1, sal, cnt ORDER BY cnt;
\end{lstlisting}

Listing~\ref{lst:client-push-up} shows a query that ranks employees according to the
fact how many salaries of other employees their own salary dominates. With {\em
    client presentation push-up}, salaries are compared in universal instead of
client format, which is correct because of the equality-preserving property
(c.f.~Corollary~\ref{def:equiv-preserving}) and saves half of the function calls
in the sub-query.

{\em Conversion push-up}, as shown in Listing~\ref{lst:conversion-push-up},
reduces the number of function calls dramatically: First, as it only converts
salaries in the end, salaries of employees aged less than 45 do not have to be
considered at all. Second, the WHERE clause converts the constant ({\tt 100K})
instead of the attribute ({\tt E\_salary}). As the outcome of conversion
functions is immutable (c.f.~\S~\ref{sec:mtsql:conv-functions}) and $C$ is also
constant, the conversion functions have to be called only once per tenant and
are then cached by the DBMS for the rest of the query execution, which becomes
much faster as we will see in Section~\ref{sec:experiments}.

\begin{lstlisting}[caption={Example for conversion push-up},
    label={lst:conversion-push-up}, language=SQL]
-- before optimization
SELECT AVG(X.sal) FROM (
  SELECT currencyFromUniversal(currencyToUniversal(E_salary, E_ttid), C) as sal
  FROM Employees WHERE E_age >= 45 AND
  currencyFromUniversal(currencyToUniversal(E_salary, E_ttid), C) > 100K) as X;
-- after optimization
SELECT AVG(currencyFromUniversal(currencyToUniversal(X.sal, X.sal_ttid),C)) FROM (
  SELECT E_salary as sal, E_ttid as sal_ttid
  FROM Employees WHERE E_age >= 45 AND
  E_salary > currencyFromUniversal(currencyToUniversal(100K, E_ttid), C) as X);
\end{lstlisting}

\subsubsection{Aggregation Distribution}

\todo[inline]{move the equations to the appendix? [done]}

Many analytical queries contain aggregation functions, some of which aggregate
on {\em convertible} attributes. The idea of aggregation distribution is to
aggregate in two steps: First, aggregate per tenant in that specific tenant
format (requires no conversion) and second, convert intermediary results to
universal (one conversion per tenant), aggregate those and convert the final
result to client format (one additional conversion). This simple idea reduces
the number of conversion function calls for $N$ records and $T$ different data
owners of these records from $(2N)$ to $(T+1)$. This is significant because $T$
is typically much smaller than $N$ (and cannot be greater).

\todo[inline]{Maybe say that this is a special form of conversion push-up
    optimization? [done --> please proof read next sentence!]}

Compared to pure {\em conversion push-up}, which works for any conversion
function pair, the applicability of {\em aggregation distribution} depends on
further algebraic properties of these functions.
Gray et al.~\cite{gray1997data} categorize
numerical aggregation functions into three categories with regard to their
ability to distribute: {\em distributive} functions, like {\tt COUNT}, {\tt
    SUM}, {\tt MIN} and {\tt MAX} distribute with functions $F$ (for partial)
and $G$ (for total aggregation). For {\tt COUNT} for instance, $F$ is {\tt
    COUNT} and $G$ is {\tt SUM} as the total count is the sum of all partial
counts. There are also {\em algebraic} aggregation functions, e.g. {\em AVG},
where the partial results are not scalar values, but tuples. In the case of {\em
    AVG}, this would be the pairs of a partial sums and partial counts because
the total average can be computed from the sum of all sums, divided by the sum
of all counts. Finally, {\em holistic} aggregation functions cannot be
distributed at all.

\begin{table}[ht]
    \centering
    \scriptsize{
    \tabcolsep 2pt
    \begin{tabular}{|l|c|c|C {1.7cm}|C {1.9cm}|}
    \hline
     & $to(x)=c \cdot x$ & $to(x)=a \cdot x + b$ & {\em to} = order-\newline preserving & {\em
         to} = equality-\newline preserving \\ \hline
     COUNT          & \tick  & \tick    & \tick  & \tick    \\ \hline
     MIN            & \tick  & \tick    & \tick  & \cross   \\ \hline
     MAX            & \tick  & \tick    & \tick  & \cross   \\ \hline
     SUM            & \tick  & \tick  & \cross & \cross   \\ \hline
     AVG            & \tick  & \tick    & \cross & \cross   \\ \hline
     {\em Holistic} & \cross & \cross   & \cross & \cross   \\ \hline
    \end{tabular}
    }
    \caption{Distributability of different aggregation functions over different
        categories of conversion functions}
    \label{tab:distributability}
\end{table}

We would like to extend the notion of Gray et al.~\cite{gray1997data} and define
the {\em distributability of an aggregation function $a$ with respect to a
    conversion function pair $(from,to)$}. Table~\ref{tab:distributability} shows
some examples for different aggregation and conversion functions. First of all,
we want to state that, as all conversion functions have scalar values as input
and output, they are always fully-{\tt COUNT}-preserving, which means that {\tt
    COUNT} can be distributed over all sorts of conversion functions. Next, we
observe that all {\em order-preserving functions} preserve the minimum and the
maximum of a given set of numbers, which is why {\tt MIN} and {\tt MAX}
distribute over the first three categories of conversion functions displayed in
Table~\ref{tab:distributability}. We further notice that if $to$ (and consequently
also $from$) is a multiplication with a constant (first column of
Table~\ref{tab:distributability}), $to$ is fully- {\tt MIN}-, fully-{\tt MAX}- and
fully-{\tt SUM}-preserving, which is why these aggregation functions distribute.
As {\tt SUM} and {\tt COUNT} distribute, {\tt AVG}, an algebraic function,
distributes as well.

Finally looking at the second column of Table~\ref{tab:distributability}, we see that
even linear functions are {\tt SUM}- and {\tt AVG}-preserving. To see why, we
can think about computing the average over all tenants as a weighted average of
partial (per-tenant) averages for {\tt AVG} and multiply these partial averages
with the partial counts to reconstruct the total sum. This method is further
explained and proven in Appendix~\ref{app:distributability}.

We conclude this subsection by observing that the conversion function pair for
{\em phone format} (c.f.~Listings~\ref{lst:pToUniv} and \ref{lst:pFromUniv}) is
not even {\em order-preserving} and does therefore not distribute while the pair
for {\em currency format} (c.f.~Listings~\ref{lst:cToUniv}\ref{lst:cFromUniv})
distributes over all standard SQL aggregation functions. An example of how this
can be used is shown in Listing~\ref{lst:distribution}.

\begin{lstlisting}[caption={Example for conversion function distribution},
    label={lst:distribution}, language=SQL]
-- before optimization
SELECT SUM(currencyFromUniversal(currencyToUniversal(E_salary, E_ttid), C)) as sum_sal FROM Employees
-- after optimization
SELECT currencyFromUniversal(SUM(t.E_partial_salary), C) as sum_sal FROM (SELECT currencyToUniversal(SUM(E_salary), E_ttid) as E_partial_salary FROM Employees GROUP BY E_ttid) as t;
\end{lstlisting}
\vspace{-5pt}

\subsubsection{Function Inlining}

As explained in \S~\ref{sec:mtsql:conv-functions}, there are several ways how to
define conversion functions. However, if they are defined as a SQL statement
(potentially including lookups into meta tables), they can be directly inlined
into the rewritten query in order to save calls to UDFs. Function inlining
typically also enables the query optimizer of the underlying DBMS to optimize
much more aggressively. In {\tt WHERE} clauses, conversion functions could
simply be inlined as sub-queries, which, however often results in sub-optimal
performance as calling a sub-query on each conversion is not much cheaper than
calling the corresponding UDF. For {\tt SELECT} clauses, the SQL standard does
anyway not allow to inline as a sub-query as this can result in attributes not
being contained neither in an aggregate function nor in the {\tt GROUP BY}
clause, which is why most commercial DBMS reject such queries (while PostgreSQL,
for instance executes them anyway). This is why the proper way to inline
functions is by using a join as shown in Listing~\ref{lst:inlining}. Our results
in Section~\ref{sec:experiments} suggest that function inlining, though
producing complex-looking SQL queries, results in very good query execution
performance.

\begin{lstlisting}[caption={Example for function inlining},
    label={lst:inlining}, language=SQL]
-- before optimization
SELECT currencyFromUniversal(currencyToUniversal(E_salary, E_ttid), C)) as E_salary FROM Employees
-- after optimization
SELECT (C1.CT_from_universal * C2.CT_to_universal * E_salary) as E_salary
FROM Employees, Tenant T1, Tenant T2, CurrencyTransform1, CurrencyTransform2
WHERE T1.T_tenant_key = C AND T1.T_currency_key = CurrencyTransform1.CT_currency_key AND
T2.T_tenant_key = E_ttid AND T2.T_currency_key = CurrencyTransform2.CT_currency_key
\end{lstlisting}
\vspace{-10pt}

\section{MT-H Benchmark}
\label{sec:mt-bench}

As MTSQL is a novel language, their exists no benchmark to evaluate the
performance on an engine that implements it, like for instance \mtbase. So far,
there exists no standard benchmark for {\em cross-tenant query processing}, only
for {\em data integration} \cite{tpcdi} which does not assume the data to be in
shared tables.
Transactions in \mtbase~are not much different from standard transactions.
Analytical queries, however, typically involve a lot of conversions and
therefore thousands of (potentially expensive) calls to UDFS. Thus, the ability
to study the usefulness of different optimizations passes on different
analytical queries was a primary design goal, which is why we decided to extend
the well-known TPC-H database benchmark~\cite{tpch}.  Our new benchmark, which
we call {\em MT-H}, extends TPC-H in the following way:

\begin{itemize}
    \item Each tenant represents a different company. The number of tenants $T$
        is a parameter of the benchmark. {\em ttids} range from $1$ to $T$.
    \item We consider {\tt Nation}, {\tt Region}, {\tt Supplier}, {\tt Part},
        and\linebreak{\tt Partsupp} common, publicly available knowledge. They are
        therefore {\em global} tables and need no modification.
    \item We consider {\tt Customer}, {\tt Orders} and {\tt Lineitem} {\em
            tenant-specific}. While the latter two are quite obviously {\em
            tenant-specific} (each company processes their own orders and line
        items), customers might actually do business with several companies.
        However, as customer information might be sensitive
        and the format of this information might differ from tenant to tenant,
        it makes sense to have specific customers per tenant.
    \item All primary keys and foreign keys relating to tenant-specific tables
        ({\tt C\_custkey}, {\tt O\_orderkey}, {\tt O\_custkey},\linebreak{\tt
            L\_orderkey}) are tenant-specific. If not mentioned otherwise, the
        attributes in {\tt Customer}, {\tt Orders} and\linebreak{\tt Lineitem}
        are {\em comparable}.
    \item We consider two domains for {\em convertible attributes} and
        corresponding functions: {\em currency} and {\em phone format}.  {\em
            currency} refers to monetary values, i.e. {\tt
            C\_acctbal},\linebreak {\tt O\_totalprice} and {\tt
            L\_extendedprice} and uses the conversion functions from
        Listings~\ref{lst:cToUniv} and \ref{lst:cFromUniv}. {\em phone format}
        is used in {\tt C\_phone} with the conversion function pair of
        Listings~\ref{lst:pToUniv} and \ref{lst:pFromUniv}. We modified the {\em
            data generator} of TPC-H (dbgen) to take the specific currency and
        phone formats into account.  Each tenant is assigned a random {\em
            currency} and {\em phone format}, except for tenant 1 who gets the
        universal format for both. 
    \item The TPC-H scaling factor $sf$ also applies to our benchmark and
        dictates the overall size of the tables. After creating all records with
        dbgen, each record in {\tt Customer}, {\tt Orders} and {\tt Lineitem}
        is assigned to a tenant in a way that foreign-key constraints are
        preserved (e.g. orders of a specific tenant link to a customer of that
        same tenant). There are two ways how this assignment happens, either
        uniform (each tenant gets the same amount of records) or zipfian (tenant
        1 gets the biggest share and tenant $T$ the smallest). This {\em tenant
            share distribution} $\rho$ is another parameter of the benchmark.
    \item We use the same 22 queries and query parameters as TPC-H.
        Additionally, for each query run, we have to define the client $C$ who runs
        the queries as well as the dataset $D$ she wants to query.
    \item For query validation, we simply set $C=1$ and $D=\{1,2,$
        \linebreak$\ldots,T\}.$ That way, we make sure to process all data and
        that the result is presented in universal format and can therefore be
        compared to expected query results of the standard TPC-H. An exception
        are queries that contain joins on {\tt O\_custkey = C\_custkey}. In
        MT-H, we make sure that each order links to a customer from the same
        tenant, thus the mapping between orders and customers is no longer the
        same as in TPC-H (where an order can potentially link to any customer).
        For such queries, we define the result from the canonical rewrite
        algorithm (without optimizations) to be the {\em gold standard} to
        validate against.
\end{itemize}

\section{Experiments and Results}
\label{sec:experiments}

\todo[inline]{Put nicer graphics... [done]}

This section presents the evaluation of \mtbase~using the MT-H benchmark. We
first evaluated the benefits of different optimization steps from
Section~\ref{sec:optimizations} and found that the combination of all of these
steps brings the biggest benefit. Second, we analyzed how \mtbase~scales with an
increasing number of tenants. With all optimizations applied and for a dataset
of 100 GB on a single machine, \mtbase~scales up to thousands of tenants with
very little overhead.  We also validated result correctness as explained in
Section~\ref{sec:mt-bench} and can report only positive results.

\subsection{Setup}

In our experiments, we used the following two setups: The first setup is a
PostgreSQL 9.6 Beta installation, running on Debian Linux 4.1.12
on a 4x16 Core AMD Opteron 6174 processor with 256 GB of main memory.
The second installation runs a commercial database (which we will call {\em
    System~C}) on a commercial operating system and on the same processor with
512 GB of main memory. Although both machines have enough secondary storage
capacity available, we decided to configure both database management systems to
use in-memory backed files in order to achieve the best performance possible.
Moreover, we configured the systems to use all available threads, which enabled
{\em intra-query parallelism}.

\begin{table*}[ht!]
    \tabcolsep 3pt
    \centering\scriptsize{
    \begin{tabular}{|l|c|c|c|c|c|c|c|c|c|c|c|c|c|c|c|c|c|c|c|c|c|c|}
    \hline
    {\bf Level} & {\bf Q01} & {\bf Q02} & {\bf Q03} & {\bf Q04} & {\bf Q05} & {\bf Q06} & {\bf Q07} & {\bf Q08} & {\bf Q09} & {\bf Q10} & {\bf Q11} & {\bf Q12} & {\bf Q13} & {\bf Q14} & {\bf Q15} & {\bf Q16} & {\bf Q17} & {\bf Q18} & {\bf Q19} & {\bf Q20} & {\bf Q21} & {\bf Q22} \\
    \hline\hline
    tpch-0.1G & 2.6 & 0.11 & 0.27 & 0.35 & 0.15 & 0.29 & 0.18 & 0.14 & 0.59 & 0.36 & 0.081 & 0.37 & 0.26 & 0.27 & 0.77 & 0.12 & 0.081 & 0.89 & 0.12 & 0.13 & 0.57 & 0.081 \\
    \hline
    canonical & 84 & 1.0 & 0.55 & 0.65 & 0.32 & 1.0 & 0.29 & 0.36 & 4.9 & 0.91 & 0.37 & 0.55 & 0.63 & 0.98 & 3.1 & 1.2 & 0.49 & 1.7 & 0.3 & 2.8 & 0.66 & 2.0 \\
    \hline
    o1 & 2.7 & 1.0 & 0.43 & 0.61 & 0.22 & 0.43 & 0.23 & 0.56 & 3.8 & 0.76 & 0.37 & 0.55 & 0.92 & 0.56 & 0.91 & 1.2 & 0.48 & 1.6 & 0.3 & 2.8 & 0.66 & 0.085 \\
    \hline
    o2 & 2.7 & 1.0 & 0.42 & 0.61 & 0.22 & 0.43 & 0.22 & 0.57 & 3.9 & 0.76 & 0.38 & 0.55 & 0.89 & 0.56 & 0.96 & 1.2 & 0.5 & 1.7 & 0.3 & 2.8 & 0.67 & 0.085 \\
    \hline
    o3 & 2.7 & 1.0 & 0.43 & 0.61 & 0.22 & 0.43 & 0.23 & 0.56 & 3.9 & 0.76 & 0.37 & 0.55 & 0.92 & 0.56 & 0.91 & 1.2 & 0.48 & 1.6 & 0.3 & 2.8 & 0.66 & 0.085 \\
    \hline
    o4 & 2.7 & 1.0 & 0.43 & 0.62 & 0.22 & 0.43 & 0.23 & 0.61 & 4.1 & 0.78 & 0.39 & 0.56 & 0.9 & 0.57 & 1.0 & 1.2 & 0.51 & 1.7 & 0.31 & 3.1 & 0.67 & 0.085 \\
    \hline
    inl-only & 2.7 & 1.0 & 0.42 & 0.65 & 0.22 & 0.43 & 0.22 & 0.57 & 3.8 & 0.76 & 0.37 & 0.55 & 0.92 & 0.56 & 0.92 & 1.2 & 0.48 & 1.6 & 0.3 & 2.8 & 0.66 & 0.085 \\
    \hline
    \end{tabular}
    }
    \captionsetup{justification=centering}
    \caption{Response times [sec] of 22 TPC-H queries for \mtbase-on-PostgreSQL
        with, $sf=1$, $T=10$, $\rho=$ uniform, $C=1$,\\\boldmath$D=\{1\}$\unboldmath, for different
        levels of optimizations, versus TPC-H with $sf=0.1$}
    \label{tab:exp:levels-d-1}
\end{table*}

\begin{table*}[ht!]
    \tabcolsep 3pt
    \centering\scriptsize{
    \begin{tabular}{|l|c|c|c|c|c|c|c|c|c|c|c|c|c|c|c|c|c|c|c|c|c|c|}
    \hline
    {\bf Level} & {\bf Q01} & {\bf Q02} & {\bf Q03} & {\bf Q04} & {\bf Q05} & {\bf Q06} & {\bf Q07} & {\bf Q08} & {\bf Q09} & {\bf Q10} & {\bf Q11} & {\bf Q12} & {\bf Q13} & {\bf Q14} & {\bf Q15} & {\bf Q16} & {\bf Q17} & {\bf Q18} & {\bf Q19} & {\bf Q20} & {\bf Q21} & {\bf Q22} \\
    \hline\hline
    tpch-0.1G & 2.6 & 0.11 & 0.27 & 0.35 & 0.15 & 0.29 & 0.18 & 0.14 & 0.59 & 0.36 & 0.081 & 0.37 & 0.26 & 0.27 & 0.77 & 0.12 & 0.081 & 0.89 & 0.12 & 0.13 & 0.57 & 0.081 \\
    \hline
    canonical & 87 & 1.0 & 0.5 & 0.6 & 0.28 & 1.0 & 0.26 & 0.37 & 4.9 & 0.89 & 0.37 & 0.56 & 0.65 & 1.0 & 3.2 & 1.2 & 0.49 & 1.6 & 0.31 & 2.8 & 0.66 & 2.0 \\
    \hline
    o1 & 87 & 1.0 & 0.5 & 0.69 & 0.33 & 1.0 & 0.27 & 0.38 & 5.2 & 0.9 & 0.39 & 0.56 & 0.92 & 1.0 & 3.1 & 1.2 & 0.51 & 1.6 & 0.32 & 3.1 & 0.68 & 2.0 \\
    \hline
    o2 & 87 & 1.0 & 0.5 & 0.61 & 0.28 & 1.0 & 0.27 & 0.38 & 5.2 & 0.9 & 0.39 & 0.57 & 0.91 & 1.0 & 3.1 & 1.2 & 0.51 & 1.6 & 0.32 & 3.1 & 0.67 & 1.3 \\
    \hline
    o3 & 32 & 1.0 & 0.45 & 0.63 & 0.28 & 0.44 & 0.24 & 0.37 & 4.3 & 0.83 & 0.38 & 0.56 & 0.91 & 1.1 & 1.9 & 1.3 & 0.51 & 1.6 & 0.32 & 3.1 & 0.67 & 1.3 \\
    \hline
    o4 & 14 & 1.0 & 0.48 & 0.62 & 0.22 & 0.44 & 0.23 & 0.57 & 3.9 & 0.93 & 0.38 & 0.56 & 0.89 & 0.73 & 1.3 & 1.2 & 0.49 & 1.6 & 0.3 & 2.8 & 0.66 & 0.27 \\
    \hline
    inl-only & 45 & 1.0 & 0.47 & 0.61 & 0.27 & 0.64 & 0.24 & 0.58 & 4.2 & 0.94 & 0.37 & 0.55 & 0.91 & 0.73 & 2.2 & 1.2 & 0.48 & 1.7 & 0.3 & 2.8 & 0.66 & 0.27 \\
    \hline
    \end{tabular}
    }
    \captionsetup{justification=centering}
    \caption{Response times [sec] of 22 TPC-H queries for \mtbase-on-PostgreSQL
        with, $sf=1$, $T=10$, $\rho=$ uniform, $C=1$,
        \\\boldmath$D=\{2\}$\unboldmath, for different
        levels of optimizations, versus TPC-H with $sf=0.1$}
    \label{tab:exp:levels-d-2}
\end{table*}

\begin{table*}[ht!]
    \tabcolsep 3pt
    \centering\scriptsize{
    \begin{tabular}{|l|c|c|c|c|c|c|c|c|c|c|c|c|c|c|c|c|c|c|c|c|c|c|}
    \hline
    {\bf Level} & {\bf Q01} & {\bf Q02} & {\bf Q03} & {\bf Q04} & {\bf Q05} & {\bf Q06} & {\bf Q07} & {\bf Q08} & {\bf Q09} & {\bf Q10} & {\bf Q11} & {\bf Q12} & {\bf Q13} & {\bf Q14} & {\bf Q15} & {\bf Q16} & {\bf Q17} & {\bf Q18} & {\bf Q19} & {\bf Q20} & {\bf Q21} & {\bf Q22} \\
    \hline\hline
    tpch-1G & 26 & 1.2 & 4.5 & 1.4 & 1.5 & 2.9 & 3.7 & 1.3 & 9.5 & 2.2 & 0.38 & 3.9 & 8.4 & 2.7 & 5.9 & 1.2 & 0.54 & 10 & 0.3 & 2.4 & 4.8 & 0.47 \\
    \hline
    canonical & 870 & 1.1 & 6.5 & 1.5 & 3.4 & 8.7 & 3.7 & 1.7 & 19 & 11 & 0.36 & 4.1 & 4.9 & 7.3 & 28 & 1.2 & 0.57 & 12 & 0.32 & 2.6 & 5.8 & 20 \\
    \hline
    o1 & 860 & 1.1 & 6.5 & 1.5 & 3.4 & 8.7 & 3.7 & 1.7 & 19 & 11 & 0.36 & 4.1 & 4.9 & 7.3 & 28 & 1.2 & 0.62 & 12 & 0.33 & 2.7 & 5.9 & 20 \\
    \hline
    o2 & 870 & 1.1 & 6.5 & 1.5 & 3.4 & 8.6 & 3.7 & 1.7 & 19 & 11 & 0.35 & 4.1 & 4.9 & 7.2 & 28 & 1.2 & 0.57 & 12 & 0.32 & 2.6 & 5.8 & 13 \\
    \hline
    o3 & 310 & 1.1 & 5.5 & 1.5 & 3.1 & 3.1 & 3.4 & 1.6 & 11 & 10 & 0.36 & 4.1 & 4.9 & 7.3 & 12 & 1.2 & 0.55 & 12 & 0.32 & 2.6 & 5.9 & 13 \\
    \hline
    o4 & 130 & 1.1 & 3.7 & 1.5 & 1.7 & 3.1 & 3.4 & 1.4 & 11 & 4.6 & 0.38 & 4.1 & 4.9 & 4.4 & 9.1 & 1.2 & 0.59 & 12 & 0.32 & 2.6 & 5.7 & 2.2 \\
    \hline
    inl-only & 450 & 1.1 & 4 & 1.6 & 1.8 & 5.1 & 3.5 & 1.4 & 14 & 4.9 & 0.39 & 4.1 & 4.8 & 4.4 & 19 & 1.2 & 0.55 & 12 & 0.32 & 2.6 & 5.8 & 2.3 \\
    \hline
    \end{tabular}
    }
    \captionsetup{justification=centering}
    \caption{Response times [sec] of 22 TPC-H queries for \mtbase-on-PostgreSQL
        with $sf=1$, $T=10$, $\rho=$ uniform, $C=1$,
        \\\boldmath$D=\{1,2,...10\}$\unboldmath, for different levels of
        optimizations, versus TPC-H with $sf=1$}
    \label{tab:exp:levels-d-all}
    \vspace{-10pt}
\end{table*}

\subsection{Workload and Methodology}

As the MT-H benchmark has a lot of parameters and in order to make things more
concrete, we worked with the following two scenarios:

{\em Scenario 1} handles the data of a business alliance of a couple of small to
mid-sized enterprises, which means there are 10 tenants with $sf=1$ and each of
them owns more or less the same amount of data ($\rho=$uniform).
%

{\em Scenario 2} is a huge database ($sf=100$) of medical records coming from
thousands of tenants, like hospitals and private practices. Some of these
institutions have vast amounts of data while others only handle a couple of
patients ($\rho$=zipf). A research institution wants to query the entire database
(D=\{1,2,...,T\}) in order to gather new insights for the development of a new
treatment. We looked at this scenario for different numbers of $T$.

In order to evaluate the overhead of {\em cross-tenant query processing} in
\mtbase~compared to single-tenant query processing, we also measured the standard
TPC-H queries with different scaling factors. When $D$ was set to all tenants,
we compared to TPC-H with the same scaling factor as MT-H. For the cases where
$D$ had only one tenant (out of ten), we compared with TPC-H with a scaling
factor ten times smaller.

Every query run was repeated three times in order to ensure stable results. We
noticed that three runs are needed for the response times to converge (within
2\%). Thus we always report the last measured response time for each query with
two significant digits.

All experiments were executed with both setups (PostgreSQL and {\em System~C}).
Whereas the major findings were the same on both systems, PostgreSQL optimizes
conversion functions (UDFs) much better by caching their results. {\em
    System~C}, on the other hand does not allow UDFs to be defined as
deterministic and hence cannot cache conversion results. This eliminates the
effect of {\em conversion push-up} when applied to comparison predicates where
we convert the constant instead of the attribute (c.f.
Listing~\ref{lst:conversion-push-up}).  This being said, the rest of this
section only reports results on PostgreSQL while we encourage the interested
reader to also consult Appendices~\ref{app:optimizations-system-C} and
\ref{app:tenant-scaling-system-C} to confirm that the main conclusions drawn
from the PostgreSQL experiments generalize. 

\begin{table}[ht!]
    \centering{
    \small
    \begin{tabular}{|L {1.5cm}|L {4.5cm}|}
        \hline
        {\bf opt level} & {\bf optimization passes} \\
        \hline \hline
        canonical & none \\
        \hline
        o1 & trivial optimizations \\
        \hline
        o2 & o1 + client presentation push-up
            \linebreak + conversion push-up \\
        \hline
        o3 & o2 + conversion function distribution \\
        \hline
        o4 & o3 + conversion function inlining \\
        \hline
        inl-only & o1 + conversion function inlining \\
        \hline
    \end{tabular}
    \caption{Different optimization levels for evaluation}
    \label{tab:opt-levels}
    }
\end{table}

\subsection{Benefit of Optimizations}

\todo[inline]{Explain in more detail what happens in Q09 (7 times slower than
    baseline) [done]}

In order to test the benefit of the different combinations of optimizations
applied, we tested {\em Scenario 1} with different optimization levels as shown
in Table~\ref{tab:opt-levels}. From {\em o1} to {\em o4} we added optimizations
incrementally, while the last optimization level ({\em inl-only}) only applied
trivial optimizations and function inlining in order to test whether the other
optimizations are useful at all.

\begin{figure*}[ht!]
    \centering
    \begin{subfigure}[b]{.32\textwidth}
        \includegraphics[width=\textwidth]{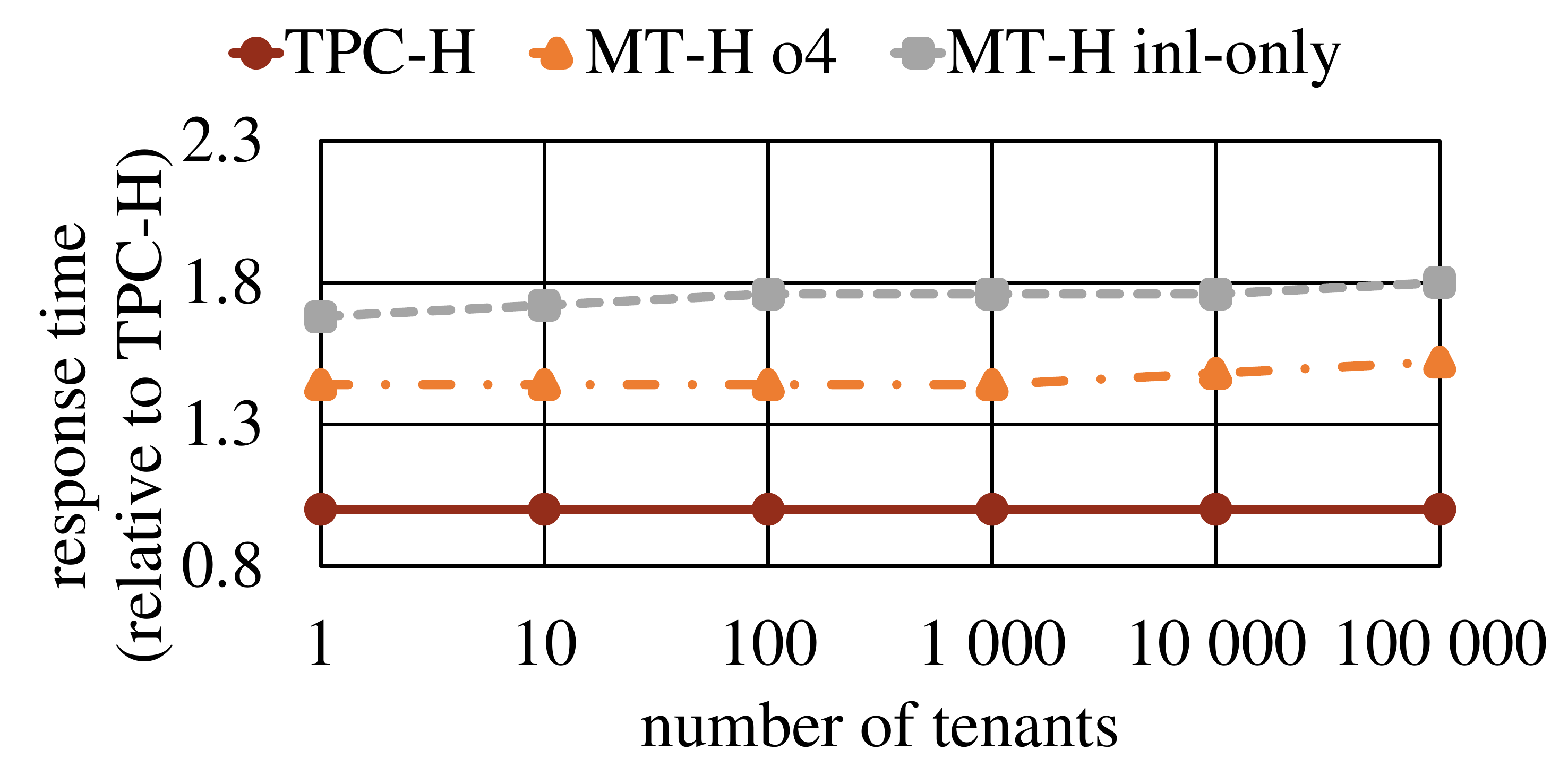}
        \caption{MT-H Query 1}
        \label{fig:exp:scaling:q1}
    \end{subfigure}
    \begin{subfigure}[b]{.32\textwidth}
        \includegraphics[width=\textwidth]{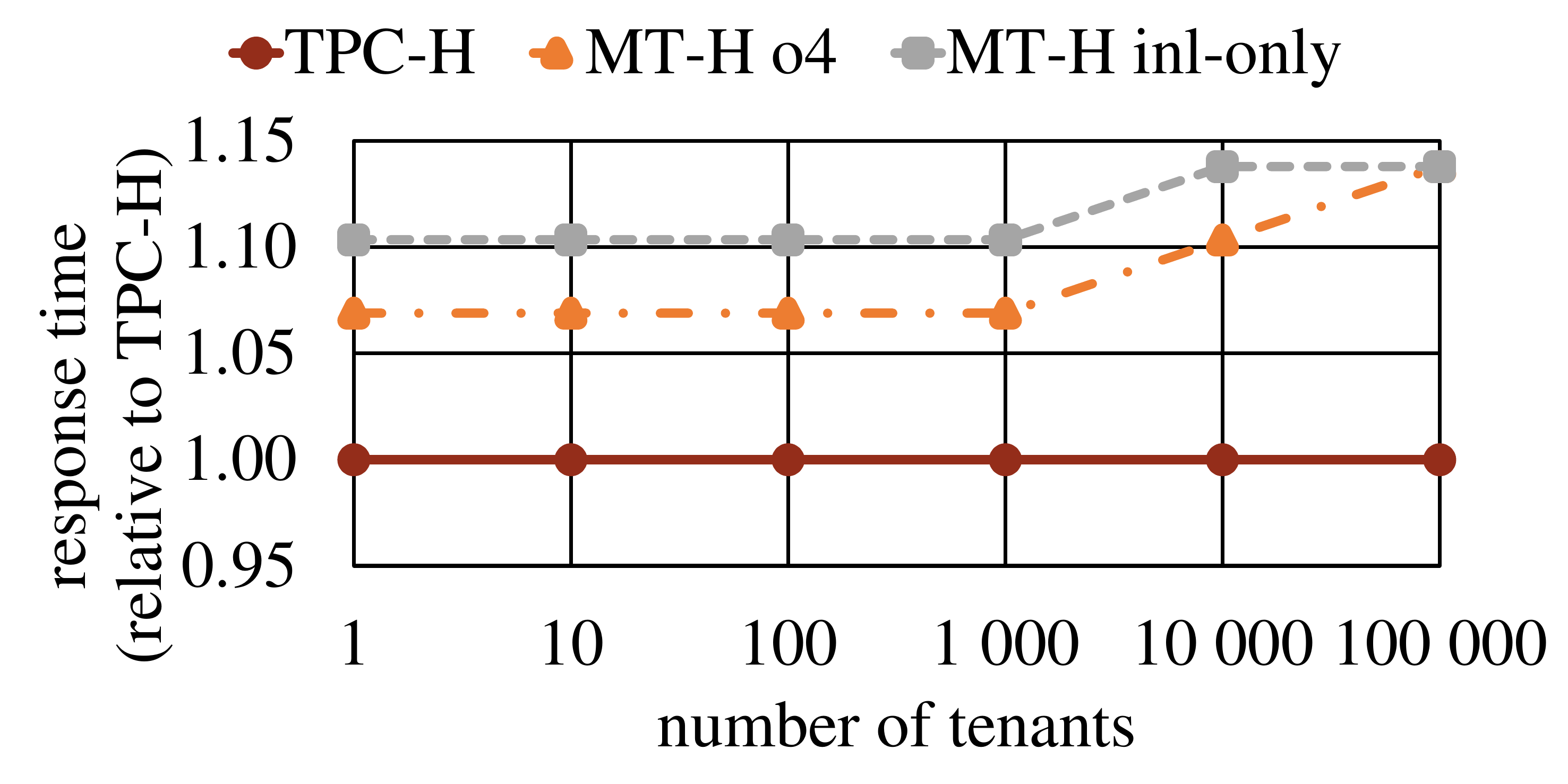}
        \caption{MT-H Query 6}
        \label{fig:exp:scaling:q6}
    \end{subfigure}
    \begin{subfigure}[b]{.32\textwidth}
        \includegraphics[width=\textwidth]{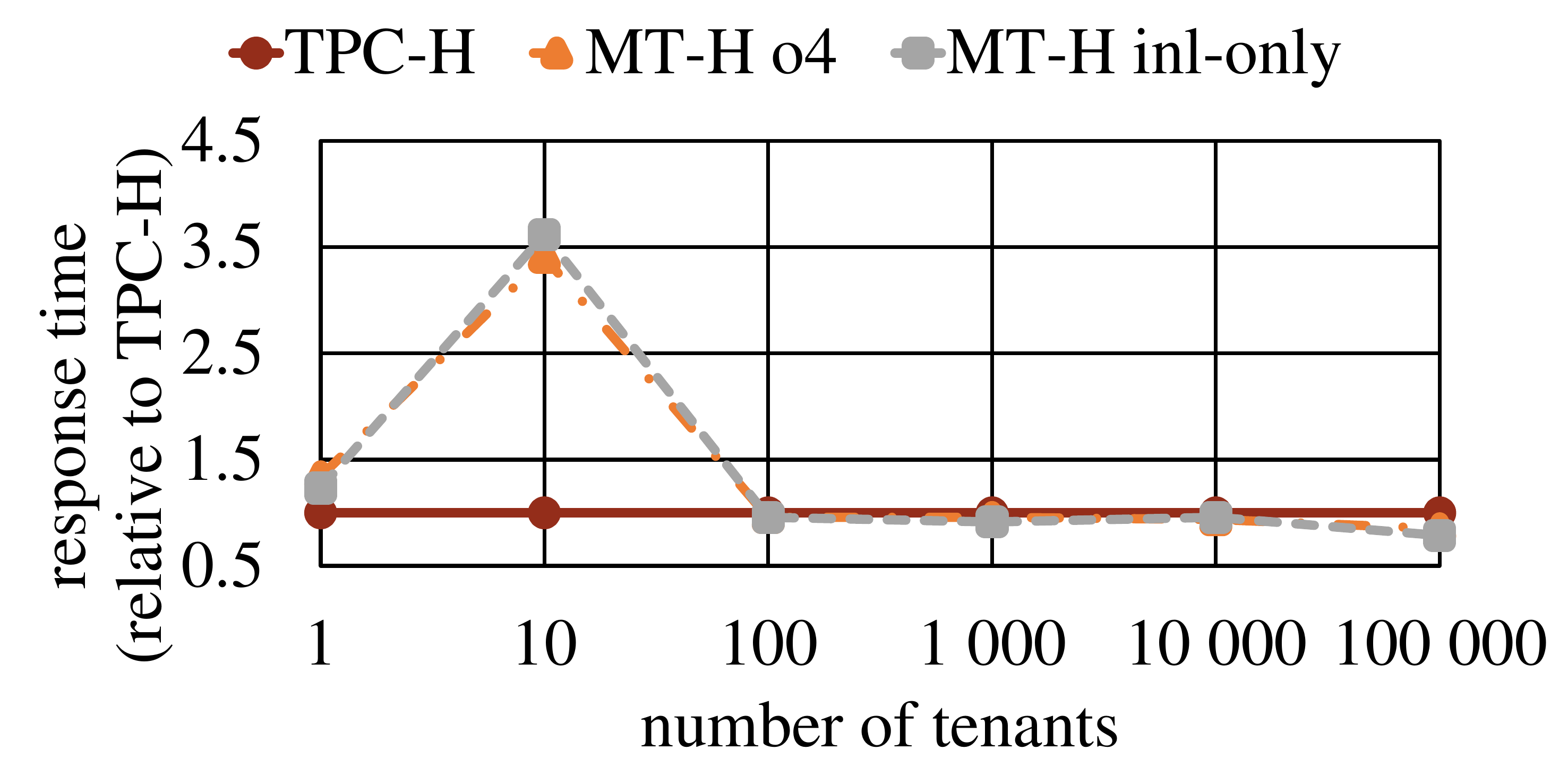}
        \caption{MT-H Query 22}
        \label{fig:exp:scaling:q22}
    \end{subfigure}
    \vspace{-5pt}
    \captionsetup{justification=centering}
    \caption{Response times (relative to TPC-H) of {\em o4} and {\em
            inlining-only} optimization levels for selected MT-H queries,
        $sf=100$,\\$T$ scaling from 1 to 100,000 on a log-scale,
        \mtbase-on-PostgreSQL}
    \label{fig:exp:scaling}
    \vspace{-5pt}
\end{figure*}

\todo[inline]{The third figure is cut off! repair!}

Table~\ref{tab:exp:levels-d-1} shows the MT-H queries for different optimization
levels and {\em Scenario 1} ($sf=1, T=10$) where client 1 queries her own data.
As we can see, in that case, applying trivial optimizations in {\em o1} is
enough because these already eliminate all conversion functions and joins and
only the D-filters remain. Executing these filters seems to be very inexpensive
because most response times of the optimized queries are close to the baseline,
TPC-H with $sf=0.1$. Queries 2, 11 and 16 however, take roughly ten times longer
than the baseline. This is not surprising when taken into account that these
queries only operate on {\em shared tables} which have ten times more data than
in TPC-H. The same effect can be observed in Q09 where a significant part of the
joined tables are shared.

Table~\ref{tab:exp:levels-d-2} shows similar results, but for $D=2$, which means that
now conversion functions can no longer be optimized away. While most of the
queries show a similar behaviour than in the previous experiment, for the ones
that involve a lot of conversion functions (i.e. queries 1, 6 and 22), we see
how the performance becomes better with each optimization pass added. We also
notice that while function inlining is very beneficial in general, it is even
more so when combined with the other optimizations.

Finally, Table~\ref{tab:exp:levels-d-all} shows the results where we query all data,
i.e. $D=\{1,2,...,10\}$. This experiment involves even more conversion functions
from all the different tenant formats into universal. In particular, when
looking again at queries 1, 6 and 22, we observe the great benefit of {\em
    conversion function distribution} (added with optimization level {\em o3}),
which, in turn, only works as great in conjunction with {\em client and
    conversion function push-up} because aggregation typically happens in the
outermost query while conversion happens in the sub-queries. Overall, {\em o4},
which contains all optimization passes that \mtbase~offers, is the clear winner.

\newpage
\subsection{Cross-Tenant Query Processing at Large}

In our final experiment, we evaluated the cost of {\em cross-tenant query
    processing} up to thousands of tenants. More concretely, we measured the
response time of conversion-intensive MT-H queries (queries 1, 6 and 22) for a
varying number of tenants between 1 and 100,000, for a large dataset where
$sf=100$ and for the best optimization level ({\em o4}) as well as for {\em
    inlining-only}. The obtained results were then compared to plain TPC-H with
$sf=100$, as shown in Figure~\ref{fig:exp:scaling}. First of all, we notice that the
cost overhead compared to {\em single-tenant query-processing} (TPC-H) stays
below a factor of 2 and in general increases very moderately with the number of
tenants.
An interesting artifact can be observed for query 22 where MT-H for one tenant
executes faster than plain TPC-H. The reason for this is a sub-optimal
optimization decision in PostgreSQL: one of the most expensive parts of query
22, namely to find customers with a specific country code, is executed with a
parallel scan in MT-H while no parallelism is used in the case of TPC-H.

\section{Related Work}
\label{sec:related-work}

\todo[inline]{adjust related work regarding DI to what we say in the introduction [done]}
\todo[inline]{remove related work on query optimization, it seems a little odd here [done]}


\mtbase~builds heavily on and extends a lot of related work. This section gives a
brief summary of the most prominent lines of work that influenced our design.
\\\\
{\bf Shared-resources (SR) systems}: In related work, this approach is also
often called {\em database virtualization} or {\em database as a service (DaaS)}
when it is used in the cloud context. Important lines of work in this domain
include (but are not limited to) {\em SqlVM}/{\em Azure SQL
    DB}~\cite{narasayya2013sqlvm,das2016daas}, {\em RelationalCloud}
\cite{mozafari2013dbseer}, {\em SAP-HANA}~\cite{sap-hana} and Snowflake
\cite{dageville2016snowflake}, most of which is well summarized in
\cite{elmore2013towards}.
{\em \mtbase} complements these systems by providing a platform that can
accommodate more, but typically smaller tenants.
\\\\
{\bf Shared-databases (SD) systems}: This approach, while appearing in the {\em
    spectrum of multi-tenant databases} by Chong et
al.~\cite{chong2006multitenancyspectrum}, is rare in practice. {\em Sql Azure
    DB} \cite{das2016daas} seems to be the only product that has an
implementation of this approach. However, even Microsoft strongly advises
against using SD and instead recommends to either use SR or ST \cite{AzureSql}.
\\\\
{\bf Shared-tables (ST) systems and schema evolution}: work in that area
includes Salesforce \cite{weissman2009salesforce}, Apache Phoenix
\cite{ApachePhoenix}, FlexScheme
\cite{aulbach2008multi,aulbach2011extensibility} and Azure SQL Database
\cite{AzureSql}. Their common idea, as in {\em MTSQL}, is to use an invisible
{\em tenant identifier} to identify which records belong to which tenant
and rewrite SQL queries in order to include filters on this {\tt ttid}. MTSQL
extends these systems by
providing the necessary features for cross-tenant query processing.
\\\\
{\bf Database Federation/Data Integration}:
The importance of {\em data integration} and its connection to MTSQL was already
stressed in the introduction.
DI is often combined with database
federation \cite{levy1998information,haas2002data}, which means that there exist
small program modules (called {\em integrators}, {\em mediators} or simply {\em
    wrappers}) to map data from different sources (possibly not all of them SQL
databases) into one common format.
While data federation generalizes well across the entire spectrum of multi-tenant
databases, maintaining such wrapper architectures is expensive, both in terms of
code maintenance and update processing. Conversely, MTSQL enables cross-tenant
query processing in a more efficient and flexible way in the context of SS and
ST databases.
\\\\
{\bf Data Warehousing}: Another approach how data integration can happen is
during extract-transform-load (ETL) operations from different (OLTP) databases
into a data warehouse \cite{kimball2002data}. Data warehouses have the
well-known drawbacks that there are costly to maintain and that the data is
possibly outdated \cite{braun2015analytics,neumann2015fast,arulraj2016bridging}.
Meanwhile, \mtbase~was specifically designed to work well in the context of
integrated OLTP/OLAP systems, also known as {\em hybrid transaction-analytical
    processing (HTAP)} systems, and could therefore be advocated as {\em
    in-situ} or {\em just-in-time} data integration.
\\\\
{\bf Security}: How to compose our proposed system with tenant data encryption
as proposed by Chong et al.~\cite{chong2006multitenancyspectrum} is not obvious
as this opens the question how tenants can process data for which they have
permission to process but which is owned by another tenant (and is therefore
encrypted with that other tenant's key). 
Obviously, simply sharing the key of tenant $t$ with
all tenants that were granted the privilege to process $t$'s data is not a
viable solution, as this allows them to impersonate $t$, which defeats the whole
purpose of encryption.
How to address this issue also depends a lot on the given {\em attacker
    scenario}: Do we want to protect tenants from each other? Do we trust the
cloud provider? Do we expect honest-but-curious behaviour or active attacks?
Also the {\em granularity} at which a tenant can share data with another tenant
matters: schema- vs.  table- vs. attribute- vs. row- vs.  predicate-based,
aggregations-only and possible combinations of these variants.  For some of
these granularities and attacker models, proposed solutions exist, e.g.
in~\cite{decapitani2007overencryption, calero2010authorizationsystems}.

\section{Conclusion}
\label{sec:conclusion}

This paper presented {\em MTSQL}, a novel paradigm to address {\em cross-tenant
    query processing} in multi-tenant databases. MTSQL extends SQL with
multi-tenancy-aware syntax and semantics, which allows to efficiently optimize
and execute cross-tenant queries in {\em \mtbase}. {\em \mtbase} is an open-source
system that implements {\em MTSQL}. At its core, it is an MTSQL-to-SQL rewrite
middleware sitting between a client and any DBMS of choice. The performance
evaluation with a benchmark adapted from TPC-H showed that {\em \mtbase} (on top
of PostgeSQL) can scale to thousands of tenants at very low overhead and that
our proposed optimizations to {\em cross-tenant queries} are highly effective.

In the future, we plan to further analyze the interplay between the {\em
    \mtbase} query optimizer and its counter-part in the DBMS execution engine in order to assess the potential of
cost-based optimizations. We also want to study conversion functions that vary over time
and investigate how MTSQL can be extended to temporal databases. Moreover, we
would like to look more into the privacy issues of multi-tenant databases,
in particular how to enable {\em cross-tenant query processing} if data is
encrypted.



\bibliographystyle{abbrv}
\bibliography{references}


\begin{appendix}

\begin{table*}[ht!]
    \tabcolsep 3pt
    \centering\scriptsize{
        \begin{tabular}{|l|c|c|c|c|c|c|c|c|c|c|c|c|c|c|c|c|c|c|c|c|c|c|}
    \hline
    {\bf Level} & {\bf Q01} & {\bf Q02} & {\bf Q03} & {\bf Q04} & {\bf Q05} & {\bf Q06} & {\bf Q07} & {\bf Q08} & {\bf Q09} & {\bf Q10} & {\bf Q11} & {\bf Q12} & {\bf Q13} & {\bf Q14} & {\bf Q15} & {\bf Q16} & {\bf Q17} & {\bf Q18} & {\bf Q19} & {\bf Q20} & {\bf Q21} & {\bf Q22} \\
    \hline\hline
    tpch-1G & 0.8 & 0.053 & 0.1 & 0.077 & 0.18 & 0.067 & 0.13 & 0.12 & 0.28 & 0.092 & 0.078 & 0.1 & 0.66 & 0.095 & 0.1 & 0.19 & 0.071 & 0.25 & 0.072 & 5.2 & 0.21 & 0.04 \\
    \hline
    canonical & 1000 & 0.12 & 230 & 0.17 & 1.7 & 8.7 & 2.6 & 3.0 & 29 & 10 & 0.3 & 0.25 & 0.66 & 8.0 & 18 & 1.3 & 2.4 & 26 & 29 & 0.099 & 0.2 & 79 \\
    \hline
    o1 & 0.78 & 0.1 & 0.23 & 0.14 & 0.087 & 0.099 & 0.95 & 0.15 & 1.1 & 0.12 & 0.29 & 0.25 & 0.65 & 0.14 & 0.14 & 1.3 & 0.13 & 8.9 & 0.91 & 0.076 & 0.19 & 0.55 \\
    \hline
    o2 & 0.77 & 0.1 & 0.23 & 0.14 & 0.087 & 0.098 & 0.97 & 0.15 & 1.0 & 0.12 & 0.28 & 0.25 & 0.66 & 0.14 & 0.14 & 1.3 & 0.13 & 8.9 & 0.94 & 0.077 & 0.2 & 3.0 \\
    \hline
    o3 & 0.78 & 0.1 & 0.23 & 0.14 & 0.088 & 0.097 & 0.93 & 0.15 & 1.1 & 0.12 & 0.28 & 0.25 & 0.65 & 0.14 & 0.14 & 1.3 & 0.14 & 8.8 & 0.92 & 0.078 & 0.2 & 3.1 \\
    \hline
    o4 & 0.78 & 0.1 & 0.3 & 0.14 & 0.089 & 0.099 & 1.0 & 0.15 & 1.0 & 0.12 & 0.29 & 0.25 & 0.66 & 0.14 & 0.14 & 1.3 & 0.14 & 8.9 & 0.92 & 0.078 & 0.2 & 0.59 \\
    \hline
    inl-only & 0.78 & 0.1 & 0.24 & 0.14 & 0.088 & 0.097 & 0.95 & 0.15 & 1.0 & 0.12 & 0.29 & 0.25 & 0.67 & 0.14 & 0.14 & 1.3 & 0.14 & 8.9 & 0.91 & 0.076 & 0.2 & 0.55 \\
    \hline
    \end{tabular}
    }
    \captionsetup{justification=centering}
    \caption{Response times [sec] of 22 TPC-H queries for \mtbase-on-System-C
        with $sf=10$, $T=10$, $\rho=$ uniform, $C=1$,
        \\\boldmath$D=\{1\}$\unboldmath, for
        different levels of optimizations, versus TPC-H with $sf=1$}
    \label{tab:app:d-1}
\end{table*}

\begin{table*}[ht!]
    \tabcolsep 3pt
    \centering\scriptsize{
        \begin{tabular}{|l|c|c|c|c|c|c|c|c|c|c|c|c|c|c|c|c|c|c|c|c|c|c|}
    \hline
    {\bf Level} & {\bf Q01} & {\bf Q02} & {\bf Q03} & {\bf Q04} & {\bf Q05} & {\bf Q06} & {\bf Q07} & {\bf Q08} & {\bf Q09} & {\bf Q10} & {\bf Q11} & {\bf Q12} & {\bf Q13} & {\bf Q14} & {\bf Q15} & {\bf Q16} & {\bf Q17} & {\bf Q18} & {\bf Q19} & {\bf Q20} & {\bf Q21} & {\bf Q22} \\
    \hline\hline
    tpch-1G & 0.8 & 0.053 & 0.1 & 0.077 & 0.18 & 0.067 & 0.13 & 0.12 & 0.28 & 0.092 & 0.078 & 0.1 & 0.66 & 0.095 & 0.1 & 0.19 & 0.071 & 0.25 & 0.072 & 5.2 & 0.21 & 0.04 \\
    \hline
    canonical & 1100 & 0.13 & 240 & 0.18 & 1.6 & 9.0 & 1.8 & 2.7 & 29 & 10 & 0.29 & 0.25 & 0.66 & 7.9 & 18 & 1.3 & 2.3 & 26 & 28 & 0.12 & 0.2 & 80 \\
    \hline
    o1 & 1100 & 0.12 & 250 & 0.16 & 1.6 & 9.0 & 1.8 & 2.9 & 30 & 11 & 0.29 & 0.25 & 0.68 & 7.9 & 18 & 1.3 & 2.4 & 26 & 29 & 0.13 & 0.19 & 80 \\
    \hline
    o2 & 1100 & 0.11 & 240 & 0.18 & 1.6 & 8.9 & 14 & 2.9 & 40 & 10 & 0.29 & 0.25 & 0.67 & 7.8 & 18 & 1.3 & 2.2 & 26 & 28 & 0.11 & 0.2 & 80 \\
    \hline
    o3 & 240 & 0.12 & 4.2 & 0.18 & 1.1 & 1.0 & 3.2 & 3.0 & 17 & 3.4 & 0.3 & 0.25 & 0.66 & 8.1 & 9.2 & 1.3 & 2.4 & 26 & 28 & 0.11 & 0.2 & 79 \\
    \hline
    o4 & 1.1 & 0.1 & 0.17 & 0.14 & 0.15 & 0.099 & 0.91 & 0.16 & 0.94 & 2.1 & 0.31 & 0.25 & 0.67 & 0.34 & 0.29 & 1.3 & 0.15 & 1.5 & 1.1 & 0.089 & 0.2 & 1.2 \\
    \hline
    inl-only & 1.7 & 0.13 & 0.2 & 0.14 & 0.15 & 0.1 & 0.78 & 0.17 & 1.1 & 0.19 & 0.28 & 0.25 & 0.61 & 0.33 & 0.21 & 1.3 & 0.15 & 1.5 & 12 & 0.099 & 0.2 & 1.1 \\
    \hline
    \end{tabular}
    }
    \captionsetup{justification=centering}
    \caption{Response times [sec] of 22 TPC-H queries for \mtbase-on-System-C
        with $sf=10$, $T=10$, $\rho=$ uniform, $C=1$, \\\boldmath$D=\{2\}$\unboldmath, for
        different levels of optimizations, versus TPC-H with $sf=1$}
    \label{tab:app:d-2}
\end{table*}

\begin{table*}[ht!]
    \tabcolsep 3pt
    \centering\scriptsize{
    \begin{tabular}{|l|c|c|c|c|c|c|c|c|c|c|c|c|c|c|c|c|c|c|c|c|c|c|}
    \hline
    {\bf Level} & {\bf Q01} & {\bf Q02} & {\bf Q03} & {\bf Q04} & {\bf Q05} & {\bf Q06} & {\bf Q07} & {\bf Q08} & {\bf Q09} & {\bf Q10} & {\bf Q11} & {\bf Q12} & {\bf Q13} & {\bf Q14} & {\bf Q15} & {\bf Q16} & {\bf Q17} & {\bf Q18} & {\bf Q19} & {\bf Q20} & {\bf Q21} & {\bf Q22} \\
    \hline\hline
    tpch-10G & 7.9 & 0.097 & 0.94 & 0.81 & 1.6 & 0.83 & 0.92 & 0.68 & 2.5 & 0.85 & 0.27 & 1.1 & 5.5 & 0.92 & 0.9 & 1.3 & 0.7 & 2.6 & 0.76 & 0.14 & 2.0 & 0.32 \\
    \hline
    canonical & 11000 & 0.14 & 2500 & 1.7 & 28 & 90 & 20 & 38 & 200 & 1100 & 0.3 & 1.2 & 6.3 & 73 & 180 & 1.3 & 2.1 & 69 & 29 & 0.17 & 3.3 & 800 \\
    \hline
    o1 & 11000 & 0.13 & 2500 & 1.6 & 28 & 90 & 21 & 37 & 190 & 1100 & 0.3 & 1.2 & 6.2 & 74 & 180 & 1.3 & 2.0 & 69 & 29 & 0.16 & 3.3 & 800 \\
    \hline
    o2 & 11000 & 0.15 & 2400 & 1.7 & 29 & 90 & 24 & 39 & 310 & 1100 & 0.3 & 1.2 & 6.2 & 74 & 180 & 1.4 & 2.1 & 69 & 30 & 0.17 & 3.4 & 790 \\
    \hline
    o3 & 2400 & 0.12 & 43 & 1.7 & 22 & 9.8 & 18 & 35 & 64 & 52 & 0.31 & 1.2 & 6.3 & 74 & 65 & 1.3 & 1.9 & 69 & 30 & 0.16 & 3.4 & 790 \\
    \hline
    o4 & 38 & 0.13 & 1.1 & 1.6 & 0.59 & 0.97 & 1.6 & 1.2 & 5.3 & 29 & 0.31 & 1.2 & 6.2 & 1.1 & 2.4 & 1.3 & 0.83 & 11 & 1.3 & 0.14 & 3.4 & 0.75 \\
    \hline
    inl-only & 42 & 0.13 & 1.8 & 1.7 & 1.6 & 1.2 & 9.8 & 1.2 & 5.5 & 13 & 0.3 & 1.2 & 6.3 & 1.1 & 1.6 & 1.3 & 0.84 & 11 & 17 & 0.18 & 3.4 & 0.53 \\
    \hline
    \end{tabular}
    }
    \captionsetup{justification=centering}
    \caption{Response times [sec] of 22 TPC-H queries for \mtbase-on-System-C
        with $sf=10$, $T=10$, $\rho=$ uniform, $C=1$, \\\boldmath$D=\{1,2,...10\}$\unboldmath, for
        different levels of optimizations, versus TPC-H with $sf=10$}
    \label{tab:app:d-all}
\end{table*}

\section{More Rewriting Examples}\label{app:more-rewriting}
While Section~\ref{sec:mtbase} of this paper focused on rewriting examples for queries,
DDL and DML statements, which generally work very similarly, were only
summarized briefly. In this paragraph we discuss some more elaborate examples.

\subsection{Rewriting DDL statements}

Section~\ref{sec:mtbase} already explained how to execute {\tt CREATE TABLE}
statements. Rewriting {\em global} constraints is also straight-forward: for
global constraints, the {\em ttids} have to be made part of the constraint. For
instance, the foreign key constraint of Listing~\ref{lst:create-table} just becomes:

{\small
\begin{verbatim}
CONSTRAINT fk_emp FOREIGN KEY (E_role_id, ttid)
  REFERENCES Roles (R_role_id, ttid)
\end{verbatim}
}

{\em Tenant-specific} check constraints are rewritten just like queries, so they
automatically include the {\em ttids} where needed. The tricky question is how
to implement {\em tenant-specific} referential integrity constraints. The way
\mtbase~implements this, is to rewrite these constraints as check constraints.
Imagine, as an example, that there is no {\em global} foreign key constraint on
the {\tt Employees} table and only tenant 0 adds this constraint privately. The
way to rewrite this to a SQL check constraint is to make sure that the set of
distinct keys in {\tt Employees\_0} is a subset of the distinct keys in {\tt
    Roles\_0}:

{\small
\begin{verbatim}
CONSTRAINT fk_emp_0 CHECK (SELECT COUNT(E_role_id)
  FROM Employees WHERE ttid=0 AND E_role_id
    NOT IN (SELECT R_role_id FROM Roles
      WHERE ttid=0)) = 0
\end{verbatim}
}

\mtbase~executes {\tt CREATE VIEW} statements by rewriting their {\tt WHERE}
clause the same way it rewrites queries (including the proper dataset, $D$). No
other modifications are needed.

\todo[inline]{The following paragraph and its example were rewritten with
the new scope definition that we have. Please proof-read!}
\subsection{Rewriting DML statements}

{\tt INSERT} statements that consist of a sub-query have to be executed in two
steps: First, the sub-query is rewritten and executed on the DBMS on behalf of
$C$. Second, for every $d \in D$, the result (which does not include any
$ttid$s) is extended with $ttid = d$ before being executed on the DBMS as a
simple {\tt INSERT} statement that contains a simple list of {\tt VALUES}.

For instance, consider tenant 0 inserting data on behalf of
tenant~1 ($C=0, D=\{1\}$) with the following statement:

{\small
\begin{verbatim}
INSERT INTO Employees VALUES E_name, E_reg_id,
    E_salary, E_age (
    SELECT E_name, E_reg_id, E_salary, E_age
    FROM Employees WHERE E_age > 40
);
\end{verbatim}
}

First of all, the intension of tenant 0 here is to copy some of his records over
to tenant 1\footnote{However, before copying data from one tenant to another,
    one should also consider the possibility to make this data global.}.  The
result of the {\tt SELECT} sub-query, executed on behalf of tenant 0 on the
exemplary database of Figure~\ref{fig:ex:running}, is {\tt ('Alice', 3, 150K, 46)}.
This record is then converted into the format of tenant 1 and extended with its
$ttid$: {\tt (1,'Alice', 3, 135K, 46)}, before being inserted into the {\tt
    Employees} table. This examples shows some of the difficulties of executing
an {\tt INSERT} (or {\tt UPDATE}) statement on behalf of somebody else. First,
as {\tt E\_emp\_id} and {\tt E\_role\_id} are {\tt NOT NULL}, they must either
have a default value or the statement fails. Second, for tenant 0 to provide a
useful {\tt E\_role\_id} for tenant 1 is difficult because it is a {\em
    tenant-specific} attribute. \mtbase~does not prevent a tenant from inserting
{\em tenant-specific} attributes, even on behalf of other tenants, but it throws
a warning in order to notify that the value might not make sense.  Luckily,
these problems do not occur with {\tt DELETE} statements.

\section{Distributability Proof}
\label{app:distributability}

In order to analyze how {\tt SUM} and {\tt AVG} distribute over linear
functions, let us formally define the multi-set of values owned by tenant $t_i$
as $X_i$ and the corresponding (linear) conversion function as $f_i(x) = a_ix + b_i$.

As we will see first, the {\em total average} over all tenants ({\tt AVG}) can
be computed as the {\em weighted average} of the {\em partial averages} of the
values of each tenant. This is proven by the following series of equations that
starts with the total average on converted values and ends up with the weighted
average of the converted partial averages:

{\scriptsize
\begin{align*}
\frac{\sum_{t \in T}\left({\sum_{x \in X_t} f_t(x)}\right)}{\sum_{t \in
        T}{|X_t|}} = \frac{\sum_{t \in T}\left({\sum_{x \in X_t}
            a_tx + b_t}\right)}{\sum_{t \in T}{|X_t|}} \\
= \frac{\sum_{t \in T}\left(a_t \left({\sum_{x \in X_t} x}\right) + b_t \cdot
        |X_t|\right)}{\sum_{t \in T}{|X_t|}} \\
= \frac{\sum_{t \in T}\left(\frac{|X_t|}{|X_t|}\left(a_t \left({\sum_{x \in
                        X_t} x}\right) + b_t \cdot |X_t|\right)\right)}{\sum_{t
        \in T}{|X_t|}} \\
= \frac{\sum_{t \in T}\left(|X_t|\left(\frac{1}{|X_t|} \cdot a_t \left({\sum_{x
                        \in X_t} x}\right) + b_t\right)\right)}{\sum_{t \in
        T}{|X_t|}} \\
= \frac{\sum_{t \in T}\left(|X_t| \cdot f \left(\frac{1}{|X_t|} \sum_{x \in
                X_t} x\right)\right)}{\sum_{t \in T}{|X_t|}}
\end{align*}
}

If we look more carefully at these equations, we realize that the term below the
fraction bar is always the same. This means that the set of equations, if we
remove this term, also shows how the total sum can be computed by multiplying
the partial averages with the corresponding partial counts, thereby proving that
also {\tt SUM} distributes over linear functions.

\begin{figure*}[ht!]
    \centering
    \begin{subfigure}[b]{0.32\textwidth}
        \includegraphics[width=\textwidth]{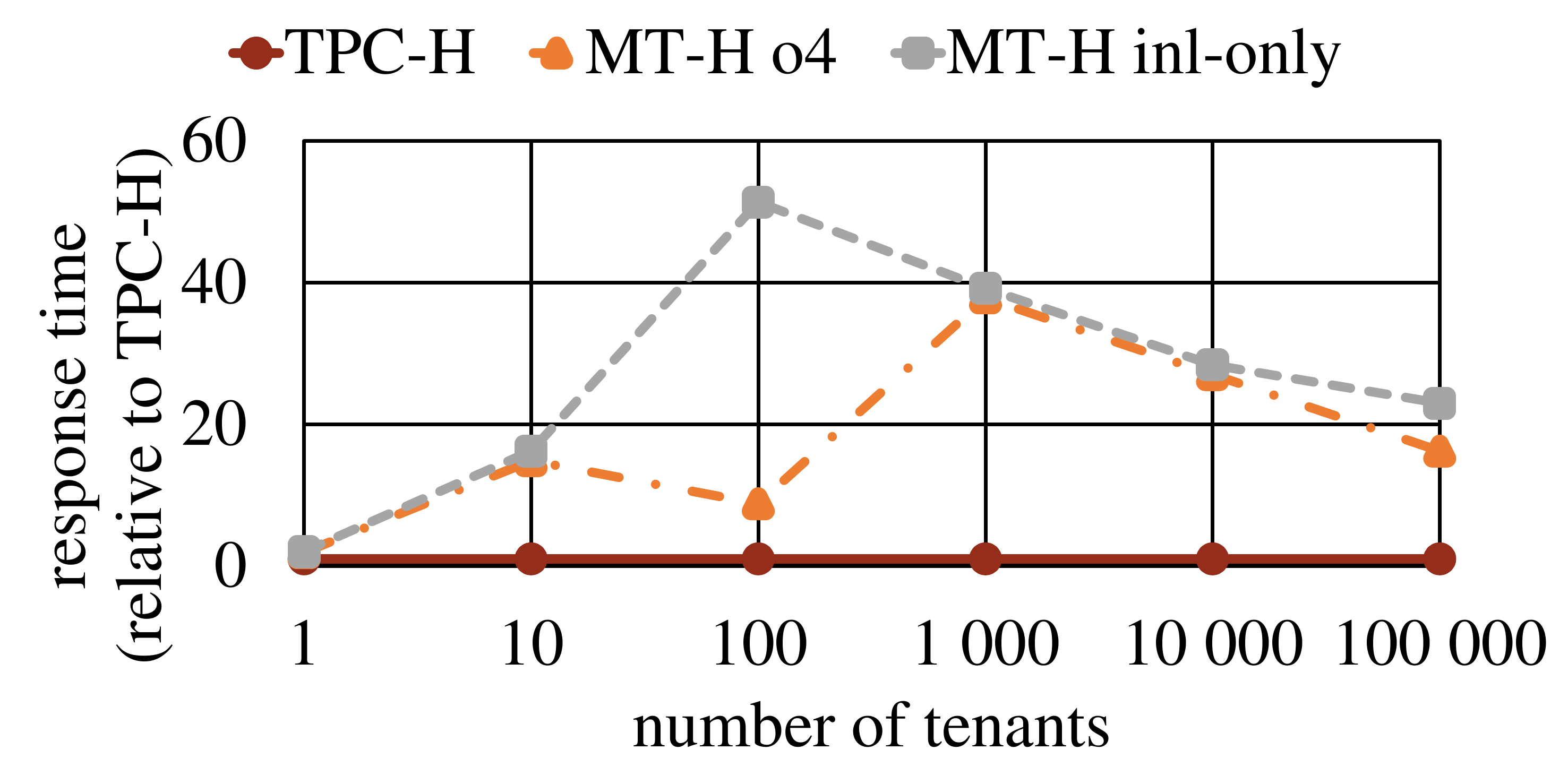}
        \caption{MT-H Query 1}
        \label{fig:exp:scaling:q1:m}
    \end{subfigure}
    \begin{subfigure}[b]{0.32\textwidth}
        \includegraphics[width=\textwidth]{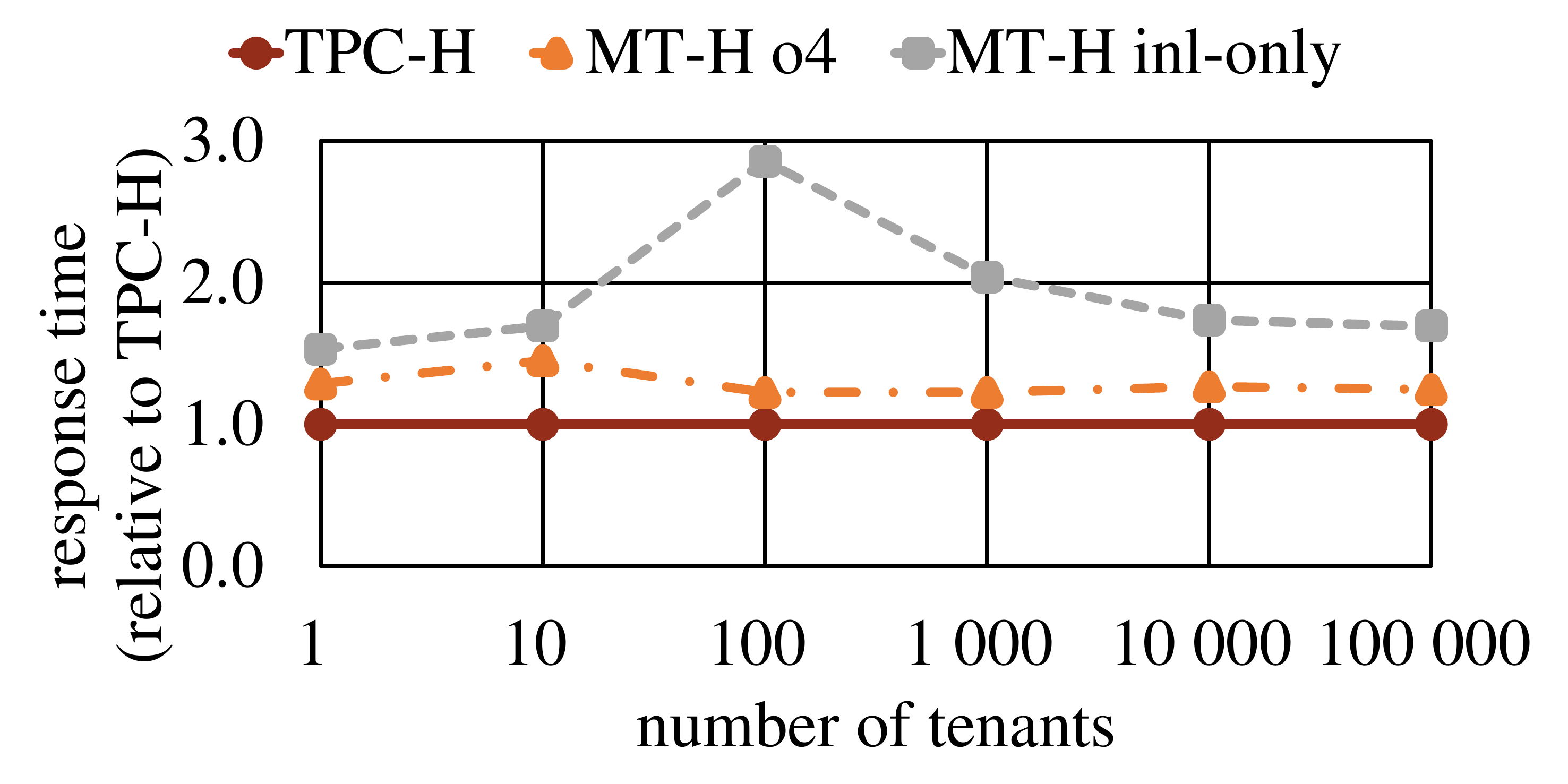}
        \caption{MT-H Query 6}
        \label{fig:exp:scaling:q6:m}
    \end{subfigure}
    \begin{subfigure}[b]{0.32\textwidth}
        \includegraphics[width=\textwidth]{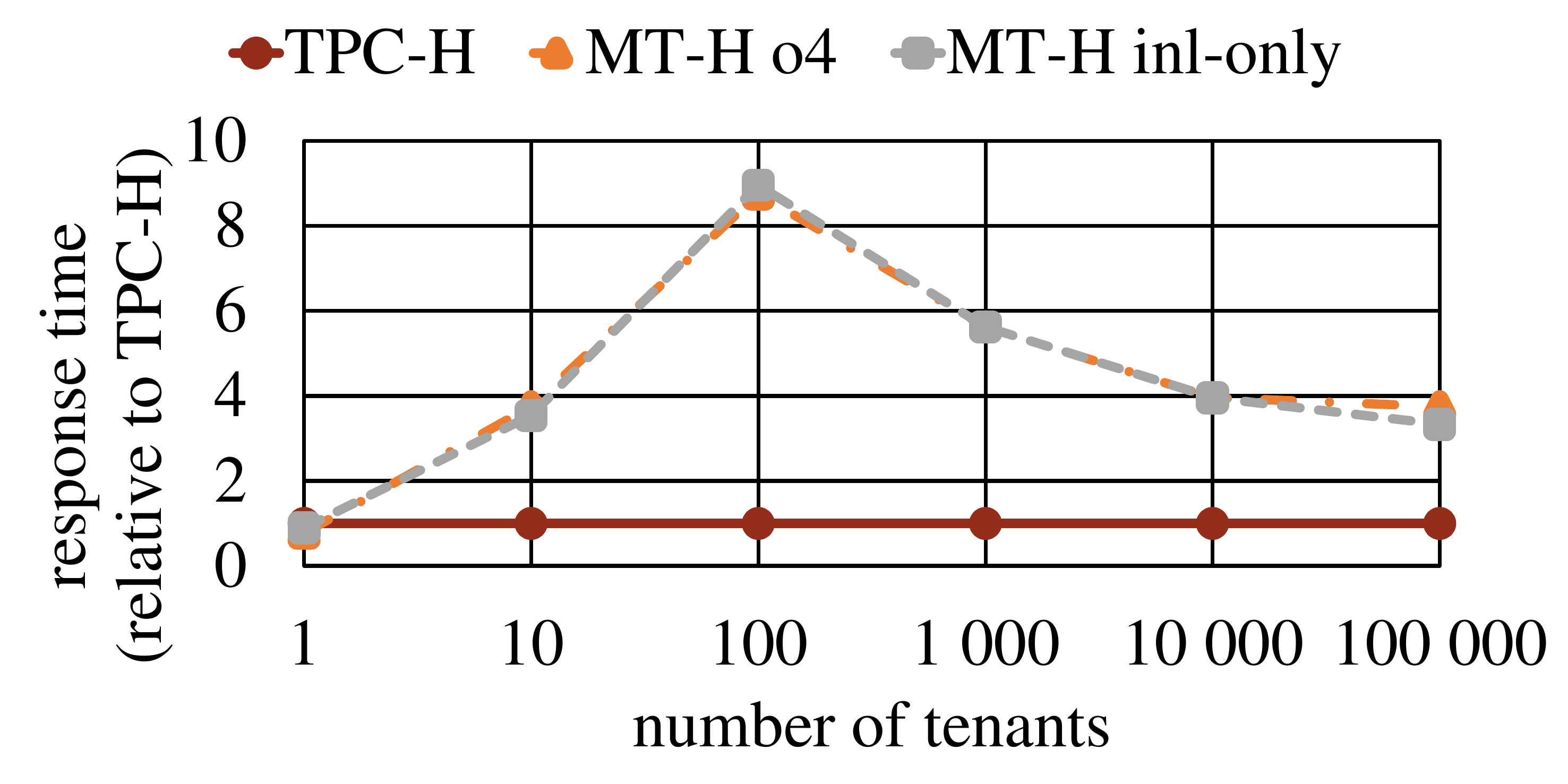}
        \caption{MT-H Query 22}
        \label{fig:exp:scaling:q22:m}
    \end{subfigure}
    \captionsetup{justification=centering}
    \caption{Response times (relative to TPC-H) of {\em o4} and {\em
            inlining-only} optimization levels for selected MT-H queries,
        $sf=100$,\\$T$ scaling from 1 to 100,000 on a log-scale,
        \mtbase-on-System-C}
    \label{fig:exp:scaling:m}
\end{figure*}

\newpage

\balance

\section{Optimizations on System~C}\label{app:optimizations-system-C}

As mentioned in Section~\ref{sec:experiments}, the performance numbers of \mtbase~on
{\em System~C} show, all in all, similar trends for optimization passes than to
the ones on PostgreSQL. The only difference is that executing conversion
functions (which are implemented as UDFs) is much more expensive in {\em System
    C} because results cannot be cached. As a consequence, executing queries
without optimizations gets much worse and Q1 can take up to three hours to be
processed on a 10\,GB dataset. The results for {\em scenario 1} are shown in
Tables~\ref{tab:app:d-1} to \ref{tab:app:d-all}.

\section{Tenant Scaling on System~C}\label{app:tenant-scaling-system-C}

An interesting picture can be seen for the tenant scaling experiment in
Figure~\ref{fig:exp:scaling:m}. While
executing the queries for a small number of tenants ($\leq 10$) or a big one
($\geq 10,000$) seems to be reasonably cheap, executing queries for a mid-sized
number of tenants seems to increase the costs dramatically. Looking at the query
plans of \emph{System~C} did not reveal much because the plans are (probably
intensionally) very coarse-grained. Thus, we can only speculate that for a
mid-sized number of tenants, the optimizer does a couple of unfortunate
decisions.

\end{appendix}

\end{document}